\documentclass[referee]{jfm}
\usepackage{natbib}

\usepackage{mathrsfs}
\usepackage{amssymb}
\usepackage{graphicx}
\usepackage{subfigure}
\usepackage{amsmath}
\usepackage{lscape}
\usepackage{setspace}

\title[Viscoelastic cross-slot flow instability]{A mechanism for oscillatory instability in viscoelastic cross-slot flow}
\author[L.~Xi \and M.~D.~Graham]{Li Xi \and Michael D.~Graham\thanks{Corresponding author. E-mail: graham@engr.wisc.edu}}
\affiliation{%
Department of Chemical and Biological Engineering\\
University of Wisconsin-Madison, Madison, WI 53706-1691
}%
\date{\today}
\begin{document}
\maketitle
\begin{abstract}

Interior stagnation point flows of viscoelastic liquids arise in a
wide variety of applications including extensional viscometry,
polymer processing and microfluidics. Experimentally, these flows
have long been known to exhibit instabilities, but the mechanisms
underlying them have not previously been elucidated. We
computationally demonstrate the existence of a supercritical
oscillatory instability of low-Reynolds number viscoelastic flow in
a two-dimensional cross-slot geometry. The fluctuations are closely
associated with the ``birefringent strand'' of highly stretched
polymer chains associated with the outflow from the stagnation point
at high Weissenberg number. Additionally, we describe the mechanism
of instability, which arises from the coupling of flow with
extensional stresses and their steep gradients in the stagnation
point region.

\end{abstract}
\section{Introduction}
While Newtonian flows become unstable only at high Reynolds number
$\textit{Re}$, when the inertial terms in momentum balance dominate,
flows of viscoelastic fluids such as polymer solutions and melts are
known to have interesting instabilities and nonlinear dynamical
behaviors even at extremely low $\textit{Re}$.
These ``purely elastic'' instabilities arise in rheometry of complex
fluids as well as in many
applications~\citep{Larson_RhA1992,Shaqfeh_ARFM1996}. Recent studies
of viscoelastic flows in microfluidic devices broaden the scope of
these nonlinear dynamical problems of viscoelastic
flows~\citep{Squires_RMP2005}. The small length scales in
microfluidic devices enable large shear rates, and thus high
$\textit{Wi}$~(Weissenberg number, $\textit{Wi}\equiv \lambda
\dot{\gamma}$, where $\lambda$ is a characteristic time scale of the
fluid and $\dot{\gamma}$ is a characteristic shear rate of the
flow),
at very low $\textit{Re}$.
Instabilities are not always undesirable, especially when the
accompanying flow modification is controllable and can thus be
utilized in the design and operation of microfluidic devices.
Specifically, instabilities have been found and flow-controlling
logic elements have been designed in a series of microfluidic
geometries, e.g. flow rectifier with anisotropic
resistance~\citep{Groisman_PRL2004}, flip-flop
memory~\citep{Groisman_Science2003} and nonlinear flow
resistance~\citep{Groisman_Science2003}. Another prospective
application of these instabilities is to enhancement of mixing at
lab-on-a-chip length scales~\citep{Groisman_Nature2001},
where turbulent mixing is absent due to
small length scales and an alternative is needed.

The best understood of these instabilities are those that occur in
viscometric flows with curved streamlines: e.g. flows in
Taylor-Couette~\citep{Muller_RhA1989},
Taylor-Dean~\citep{Joo_JFM1994},
cone-and-plate~\citep{Magda_JNNFM1988} and
parallel-plates~\citep{Groisman_Nature2000,Magda_JNNFM1988} flow
geometries. In these geometries, the primary source of instability
is the coupling of normal stresses with streamline curvature (i.e.
the presence of ``hoop stresses''), leading to radial compressive
forces that can drive
instabilities~\citep{Shaqfeh_ARFM1996,Muller_RhA1989,Joo_JFM1994,Magda_JNNFM1988,Larson_JFM1990,Pakdel_PRL1996,Graham_JFM1998}.
Similar mechanisms drive instabilities in viscoelastic free surface
flows~\citep{Speigelberg_JNNFM1996,Graham_PoF2003}.

Attention in this paper focuses on a different class of flows, whose
instabilities are not well-understood -- stagnation point flows,
like those generated with
opposed-jet~\citep{Chow_Macromole1988,Muller_JNNFM1988},
cross-slot~\citep{Arratia_PRL2006}, two-roll mill~\citep{Ng_JRh1993}
and four-roll mill~\citep{Ng_JRh1993,Broadbent_JNNFM1978} devices.
Figure~\ref{channel_geo} shows a schematic of a cross-slot geometry.
A characteristic phenomenon in these stagnation point flows is the
formation of a narrow region of fluid with high polymer stress
extending downstream from the stagnation point. This region can be
observed in optical experiments as a bright birefringent ``strand''
with the rest of fluid dark~\citep{Harlen_JNNFM1990}. Keller and
coworkers~\citep{Chow_Macromole1988,Muller_JNNFM1988} reported
instabilities in stagnation point flows of semi-dilute polymer
solutions generated by an axisymmetric opposed-jet device.
Specifically, for a fixed polymer species and concentration, upon a
critical extension rate (or critical $\textit{Wi}$) polymer chains
become stretched by flow near the stagnation point and a sharp
uniform birefringent stand forms. The width of this birefringent
strand increases with increasing $\textit{Wi}$ until a stability
limit is reached, beyond which the birefringent strand becomes
destabilized and changes in its morphology are observed. At higher
$\textit{Wi}$, the flow pattern and birefringent strand become
time-dependent. Recent tracer and particle-tracking experiments of
stagnation point flow in a micro-fabricated cross-slot geometry by
Arratia~\textit{et~al.}~\citep{Arratia_PRL2006} show instabilities
of dilute polymer solution at low $\textit{Re}$~($<10^{-2}$). In
their experiments fluid from one of the two incoming channels is
dyed and a sharp and flat interface between dyed and undyed fluids
is observed at low $\textit{Wi}$. Upon an onset value of
$\textit{Wi}$, this flow pattern loses its stability: spatial
symmetry is broken but the flow remains steady. The interface
becomes distorted in such a way that more than half of the dyed
fluid goes to one of the outgoing channels while more undyed fluid
travels through the other. At even higher $\textit{Wi}$ the flow
becomes time-dependent and the direction of asymmetry flips between
two outgoing channels with time. Particle-tracking images in the
time-dependent flow pattern indicate the existence of vortical
structures around stagnation point.

Another class of stagnation point flows is associated with
liquid-solid or liquid-gas interfaces, such as flows passing
submerged solid obstacles, around moving bubbles or toward a free
surface. For example,
McKinley~\textit{et~al.}~\citep{McKinley_PTRSLA1993} reported a
three-dimensional steady cellular disturbances in the wake of a
cylinder submerged in a viscoelastic fluid. Around a falling sphere
in viscoelastic fluids, fore-and-aft symmetry of velocity field is
broken and the velocity perturbation in the wake can be away from
the sphere, toward the sphere or a combination of the two depending
on the polymer
solution~\citep{Hassager_Nature1979,Bisgaard_RhA1982,Bisgaard_JNNFM1983}.

Remmelgas~\textit{et~al.}~\citep{Remmelgas_JNNFM1999}
computationally studied the stagnation point flow in a cross-slot
geometry with two different FENE~(finite extensible nonlinear
elastic) dumbbell models.
Using the two models, they studied the effects of
configuration-dependent friction coefficient on polymer relaxation
and the shape of the birefringent strand. Their simulation approach
was restricted to relatively low $\textit{Wi}$~($\sim \mathcal
{O}(1)$) with symmetry imposed on centerlines of all channels.
Harlen~\citep{Harlen_JNNFM2002} conducted simulations of a
sedimenting sphere in a viscoelastic fluid to explore the wake
behaviors. He explains the experimental observations of both
negative (velocity perturbation away from the sphere) and extended
(velocity perturbation toward the sphere) wakes in terms of combined
effects of the stretched polymer in the birefringent strand
following the stagnation point behind the sphere and the recoil
outside of the strand. Neither of these analyses directly addressed
instabilities of these flows.

Various approximate approaches have been taken in the past to obtain
an understanding of these instabilities observed in experiments.
Harris and Rallison~\citep{Harris_JNNFM1993,Harris_JNNFM1994}
investigated the instabilities of the birefringent strand behind a
free isolated stagnating point through a simplified approach, in
which polymer molecules are modeled as linear-locked dumbbells,
which are fully stretched within a thin strand lying along the
centerline. Polymer molecules contribute a normal stress
proportional to the extension rate only when they are fully
stretched~(i.e. in the strand), otherwise the flow is treated as
Newtonian. The lubrication approximation is applied for the
Newtonian region and the effects of birefringent strand are coupled
into the problem through point forces along the strand. Two
instabilities are reported. At low $\textit{Wi}$~($\approx1.2-1.7$),
a varicose disturbance is linearly unstable, in which the width of
birefringent strand oscillates without breaking the symmetry of the
flow pattern. At higher $\textit{Wi}$ another instability is
observed in which symmetry with respect to the extension axis breaks
and the birefringent strand becomes sinuous in shape and oscillatory
with time, with zero displacement at the stagnation point and
increasing magnitude of displacement downstream from it. Symmetry
with respect to the inflow axis is always imposed. The mechanism of
these instabilities is explained: perturbations in the shape or
position of the birefringent strand affect the stretching of
incoming polymer molecules such that they enhance the perturbation
after they become fully stretched and merge into the strand. This
mechanism is close to the one we are about to present later in this
paper with regard to the importance of flow kinematics and the
extensional stress. However, in their linear stability analysis the
spatial dependence of the birefringent strand in the outflow
direction is neglected, which is important according to our
simulations.
\"{O}ztekin~\textit{et~al.}~\citep{Oztekin_JNNFM1997} studied steady
state similarity solution for planar stagnation point flow at a
solid wall predicting that this flow is linearly unstable to local
three-dimensional disturbances. Their results indicate that
traveling wave disturbances that have
periodic structure in the neutral direction could lead to
instabilities of steady state solutions above certain critical
$\textit{Wi}$.


In this paper, we present numerical simulation results of
viscoelastic stagnation point flow in a two-dimensional cross-slot
geometry.
With increasing $\textit{Wi}$, we observe the formation and
elongation of the birefringent strand across the stagnation point.
At high $\textit{Wi}$, we find the occurrence of an oscillatory
instability.
These results resemble the experimental observations of oscillatory
birefringent width by
M\"{u}ller~\textit{et~al.}~\citep{Muller_JNNFM1988} and the varicose
instability predicted by
Harris~\textit{et~al.}~\citep{Harris_JNNFM1994}. By analyzing the
perturbations in both velocity and stress fields, a novel
instability mechanism based on normal stress effects and flow
kinematics is identified.

\section{Formulation and Methods}
\begin{figure}
\centering
\includegraphics[width=0.65\textwidth, bb=426 0 2016 1584]{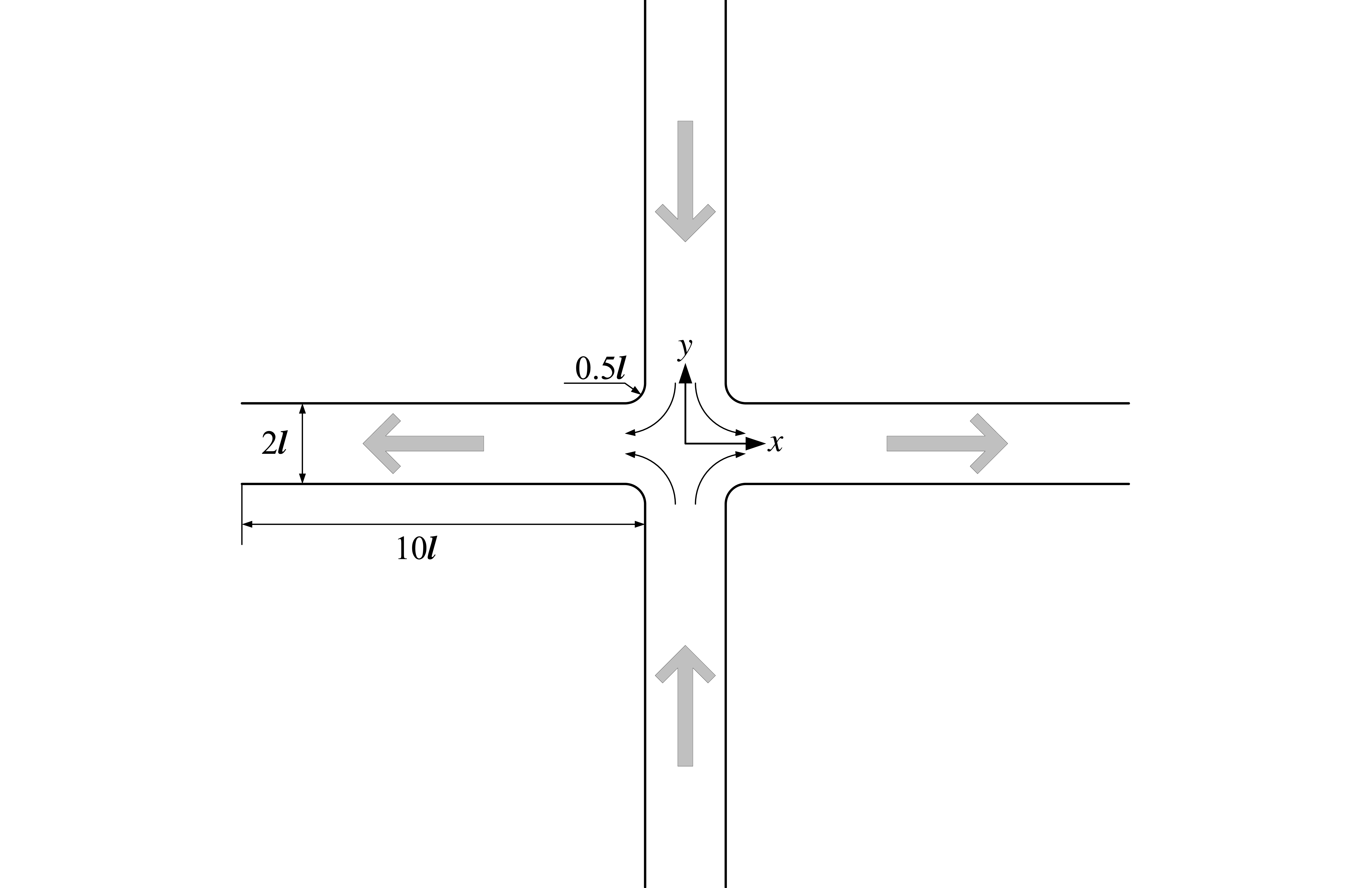}
\caption{Schematic of the cross-slot flow geometry.}
\label{channel_geo}
\end{figure}

We consider a fourfold symmetric planar cross-slot geometry, as
shown in Figure~\ref{channel_geo}. Flow enters from top and bottom
and leaves from left and right. For laminar Newtonian flow, two
incoming streams meet at the intersection of the cross and each of
them splits evenly and goes into both outgoing channels, generating
a stagnation point at the origin near which an extensional flow
exists. We use round corners at the intersections of channel walls
in order to avoid enormous stress gradients at the corners, which
cause numerical difficulties.

The momentum and mass balances are:
\begin{eqnarray}%
    \label{ns_momentum_eq}%
        \textit{Re} \left( \frac{\partial \boldsymbol{u}}{\partial t} +
        \boldsymbol{u} \cdot \boldsymbol{\nabla u} \right)= -
        \boldsymbol{\nabla}p + \beta \nabla^{2}\boldsymbol{u} + \left(1 -
        \beta\right)\frac{2}{\textit{Wi}}\left(\boldsymbol{\nabla} \cdot
        \boldsymbol{\tau}_{p}\right),%
    \\%
    \label{ns_continuity_eq}%
        \boldsymbol{\nabla} \cdot \boldsymbol{u} = 0.%
\end{eqnarray}%
Parameters in Equations~(\ref{ns_momentum_eq}) and
(\ref{ns_continuity_eq}) are defined as: $\textit{Re} \equiv \rho U
l / \left(\eta _{s} + \eta _{p}\right)$, $\textit{Wi} \equiv 2
\lambda U /l$ and $\beta \equiv \eta _{s} / \left(\eta _{s} + \eta
_{p}\right)$, where $\rho$ is the fluid density and for dilute
polymer solution we assume it to be the same as the solvent density;
$\eta _{s}$ is the solvent viscosity and $\eta _{p}$ is the polymer
contribution to the shear viscosity at zero shear rate
and $U$ and $l$ are characteristic velocity and length scales of the
flow. Here $l$ is chosen to be the half channel width and the
definition of $U$ is based on the pressure drop applied between the
entrances and exits of the channel. Specifically, $U$ is defined to
be the centerline velocity of a Newtonian plane Poiseuille flow
under the same pressure drop in a straight channel with length
$20l$, which is comparable to the lengths of streamlines in the
present geometry. According to this definition, the nondimensional
pressure drop in our simulation is fixed at $40$ and the centerline
Newtonian velocity in cross-slot geometry is typically slightly
lower than $1$ since the extensional flow near the stagnation point
has a higher resistance than that in a straight channel.
The polymer contribution to the stress tensor is denoted
$\boldsymbol{\tau}_{p}$ and is calculated with the FENE-P
constitutive equations~\citep{Bird_1987}:
\begin{eqnarray}%
    \label{fenep_conformation_eq}
        \frac{\boldsymbol{\alpha}}{1 -
        \frac{\textrm{tr}(\boldsymbol{\alpha})}{b}} + \frac{\textit{Wi}}{2} \left(
        \frac{\partial \boldsymbol{\alpha}}{\partial t} +
        \boldsymbol{u} \cdot \boldsymbol{\nabla \alpha} -
        \boldsymbol{\alpha} \cdot \boldsymbol{\nabla u} - \left( \boldsymbol{\alpha} \cdot \boldsymbol{\nabla
        u} \right)^{T} \right) = \left( \frac{b}{b + 2} \right)
        \boldsymbol{\delta},%
    \\%
    \label{fenep_stress_eq}%
        \boldsymbol{\tau}_{p} = \frac{b + 5}{b} \left( \frac{\boldsymbol{\alpha}}{1 -
        \frac{\textrm{tr}(\boldsymbol{\alpha})}{b}} -\left( 1 - \frac{2}{b + 2} \right) \boldsymbol{\delta}
        \right).%
\end{eqnarray}
In Equations~(\ref{fenep_conformation_eq}) and
(\ref{fenep_stress_eq}), polymer chains are modeled as FENE
dumbbells (two beads connected by a
finitely-extensible-nonlinear-elastic spring). Here
$\boldsymbol{\alpha} \equiv \langle \boldsymbol{QQ} \rangle$ is the
conformation tensor of the dumbbells where $\boldsymbol{Q}$ is the
end-to-end vector of the dumbbells and $\langle \cdot \rangle$
represents an ensemble average. The parameter $b$ determines the
maximum extension of dumbbells, i.e. the upper limit of
$\textrm{tr}(\boldsymbol{\alpha})$.

At the entrances and exits of the flow geometry, normal flow
boundary conditions are applied, i.e. $\boldsymbol{t} \cdot
\boldsymbol{u} = 0$ where $\boldsymbol{t}$ is the unit vector
tangential to the boundary. Pressure is set to be $40$ at entrances
and $0$ at exits. No-slip boundary conditions are applied at all
other boundaries. Boundary conditions for stress are only needed at
the entrances, where the profile of $\boldsymbol{\alpha}$ is set to
be the same as that for a fully developed pressure-driven flow in a
straight channel with the same $\textit{Wi}$. Other fixed parameters
in our simulations are: $\textit{Re} = 0.1$, $\beta = 0.95$ and $b =
1000$, which means we focus on dilute solutions of long-chain
polymers at low Reynolds number.

The discrete elastic stress splitting~(DEVSS)
formulation~\citep{Baaijens_JNNFM1997,Baaijens_JNNFM1998} is applied
in our simulation: i.e. a new variable $\boldsymbol{\Lambda}$ is
introduced as the rate of strain and a new equation is added into
the equation system:
\begin{eqnarray}%
    \label{devss_lambda_eq}
        \boldsymbol{\Lambda} = \boldsymbol{\nabla u} + \boldsymbol{\nabla
        u}^{T}.%
\end{eqnarray}
A numerical stabilization term $\gamma \boldsymbol{\nabla} \cdot
\left( \boldsymbol{\nabla u} + \boldsymbol{\nabla u}^{T} -
\boldsymbol{\Lambda}\right)$ is added to the right-hand-side of the
momentum balance~(Equation~(\ref{ns_momentum_eq})) and it is
worthwhile to point out that this term is only nontrivial in the
discretized formulation and does not change the physical problem. In
this term, $\gamma$ is an adjustable parameter and $\gamma = 1.0$ is
used in our simulations. The velocity field $\boldsymbol{u}$ is
interpolated with quadratic elements while pressure $p$, polymer
conformation tensor $\boldsymbol{\alpha}$ and rate of strain
$\boldsymbol{\Lambda}$ are interpolated with linear elements.
Consistent with Baaijens's conclusion~\citep{Baaijens_JNNFM1998},
DEVSS greatly increases the upper limit of $\textit{Wi}$ achievable
in our simulations. Quadrilateral elements are used for all
variables. Our experience shows that quadrilateral elements have
great advantages over triangular ones, yielding much better spatial
smoothness in the stress field at comparable degrees of freedom to
be solved. Another merit of quadrilateral elements is the capability
of manual control over mesh grids. This is extremely important when
certain restrictions, such as symmetry, are required. In our
simulation, finer meshes are used within and around the intersection
region of the geometry and the mesh is required to be symmetric with
respect to both axes. Within a horizontal band ($-0.2<y<0.2$) across
the stagnation point, very fine meshes are generated to capture the
sharp stress gradient along the birefringent strand. The streamline
upwind/Petrov-Galerkin method (SUPG)~\citep{Brooks_CMAME1982} is
applied in Equation~(\ref{fenep_conformation_eq}) by replacing the
usual Galerkin weighting function $w$ with $w + \delta h
\boldsymbol{u} \cdot
\boldsymbol{\nabla}w/\lVert\boldsymbol{u}\rVert$, where $h$ is the
geometric average of the local mesh length scales and $\delta$ is an
adjustable parameter, set to $\delta = 0.3$ in our simulations. This
formulation is implemented using the commercially available
\textit{COMSOL Multiphysics} software.

\section{Results and Discussions}
\subsection{Steady States}
\begin{figure}
    \centering%
    \subfigure[~$\lVert\boldsymbol{u}\rVert$]
    {%
        \label{vel_ss_2006-10-12a}%
        \includegraphics[width=0.80\textwidth]{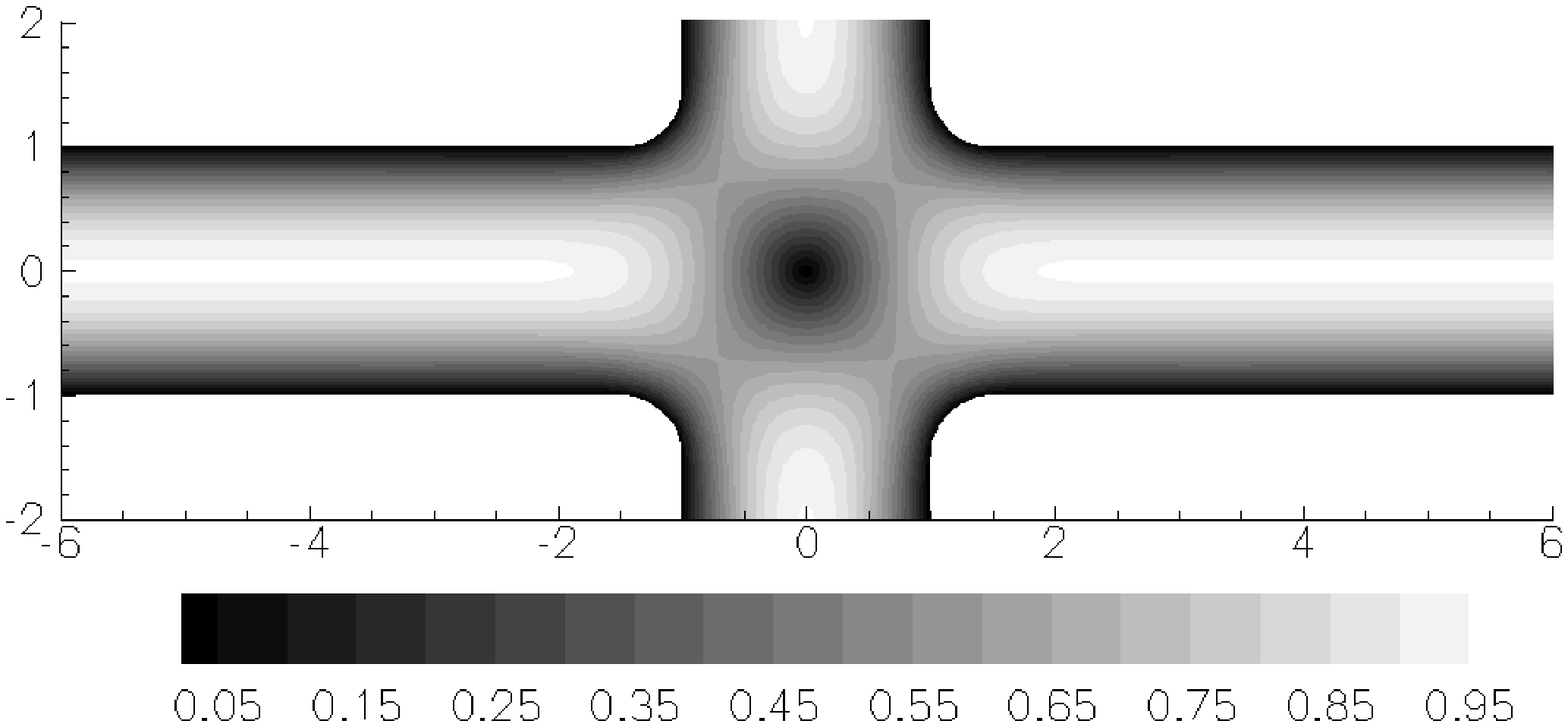}%
    }%
    \vspace{2pt}
    \\
    \subfigure[~$\partial u_{x}/\partial x$]
    {%
        \label{u1x_ss_2006-10-12a}%
        \includegraphics[width=0.80\textwidth]{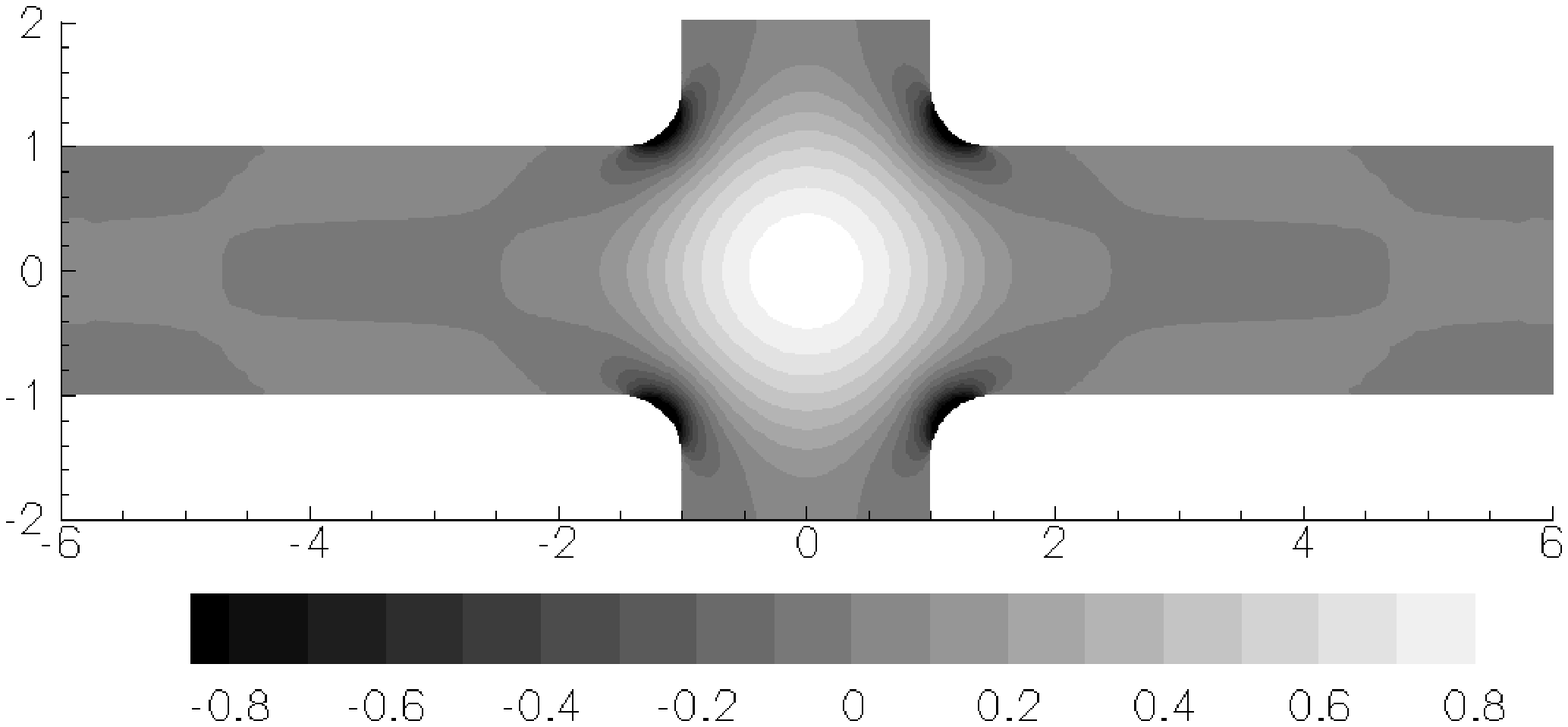}%
    }%
    \vspace{2pt}%
    \\%
    \subfigure[~$\textrm{tr}(\boldsymbol{\alpha})$]%
    {%
        \label{tra_ss_2006-10-12a}%
        \includegraphics[width=0.80\textwidth]{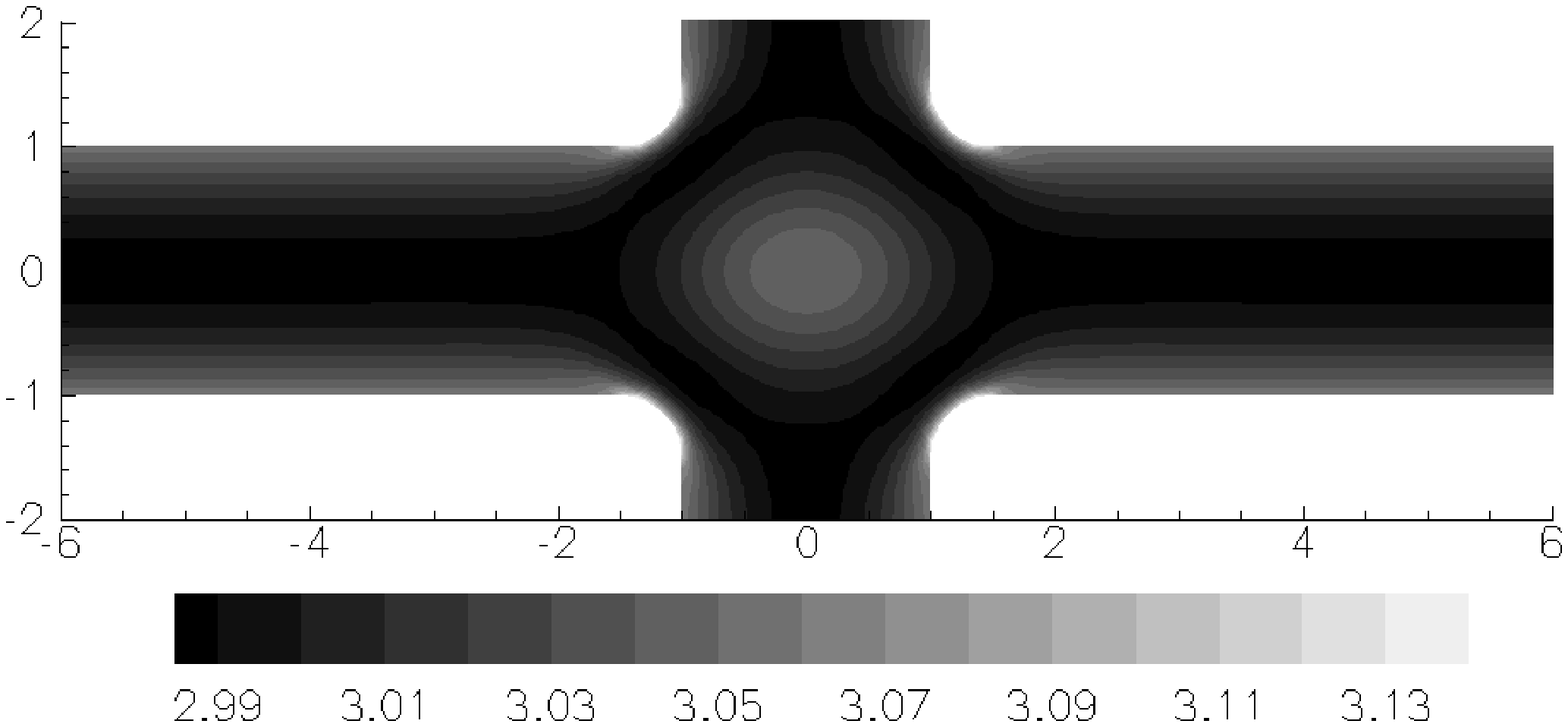}%
    }%
    \caption{Contour plots of steady state solution: $\textit{Wi} = 0.2$~(only the central part of the flow domain is shown).}%
    \label{ss_2006-10-12a}
\end{figure}

\begin{figure}
    \centering%
    \subfigure[~$\lVert\boldsymbol{u}\rVert$]%
    {%
        \label{vel_ss_2006-10-12g}%
        \includegraphics[width=0.80\textwidth]{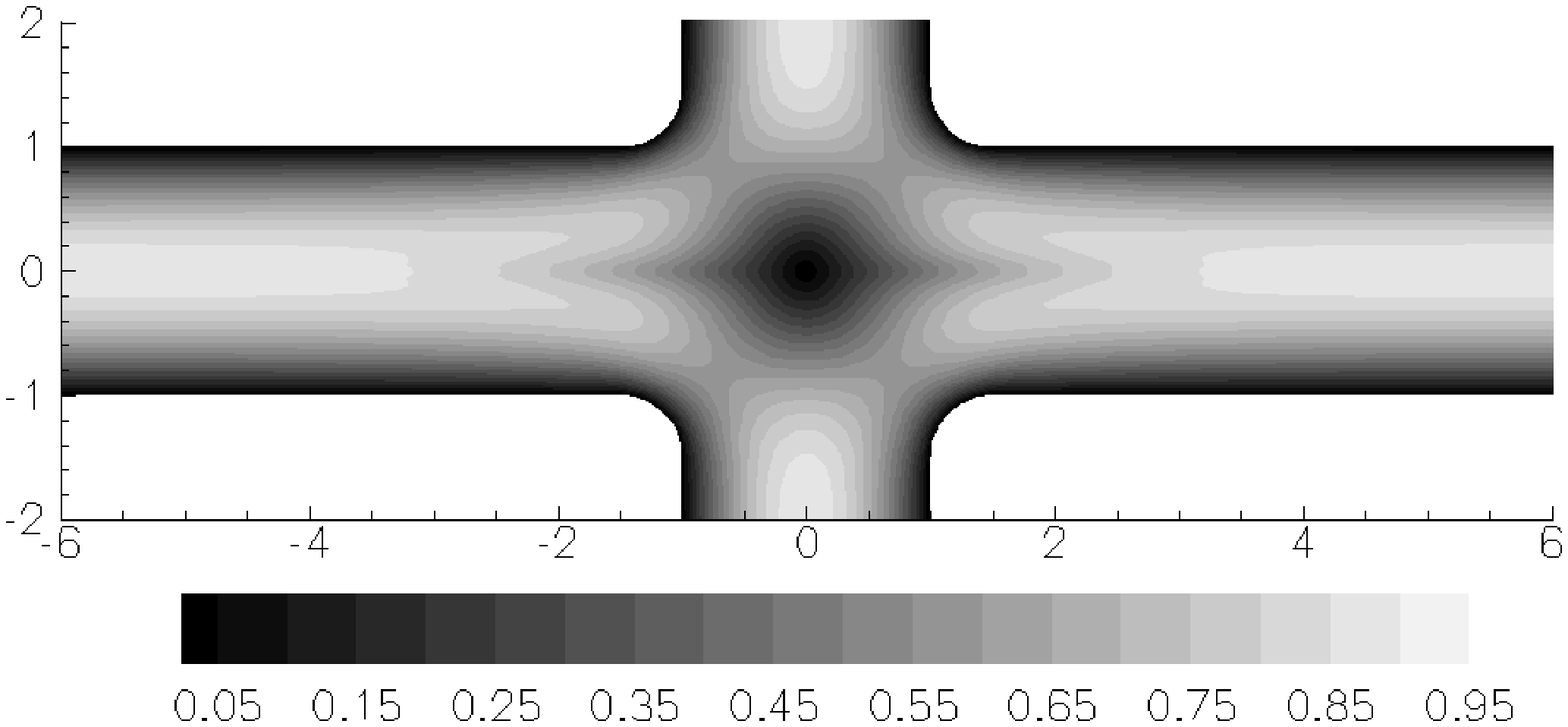}%
    }%
    \vspace{2pt}
    \\
    \subfigure[~$\dot{\epsilon} = \partial u_{x}/\partial x$]%
    {%
        \label{u1x_ss_2006-10-12g}%
        \includegraphics[width=0.80\textwidth]{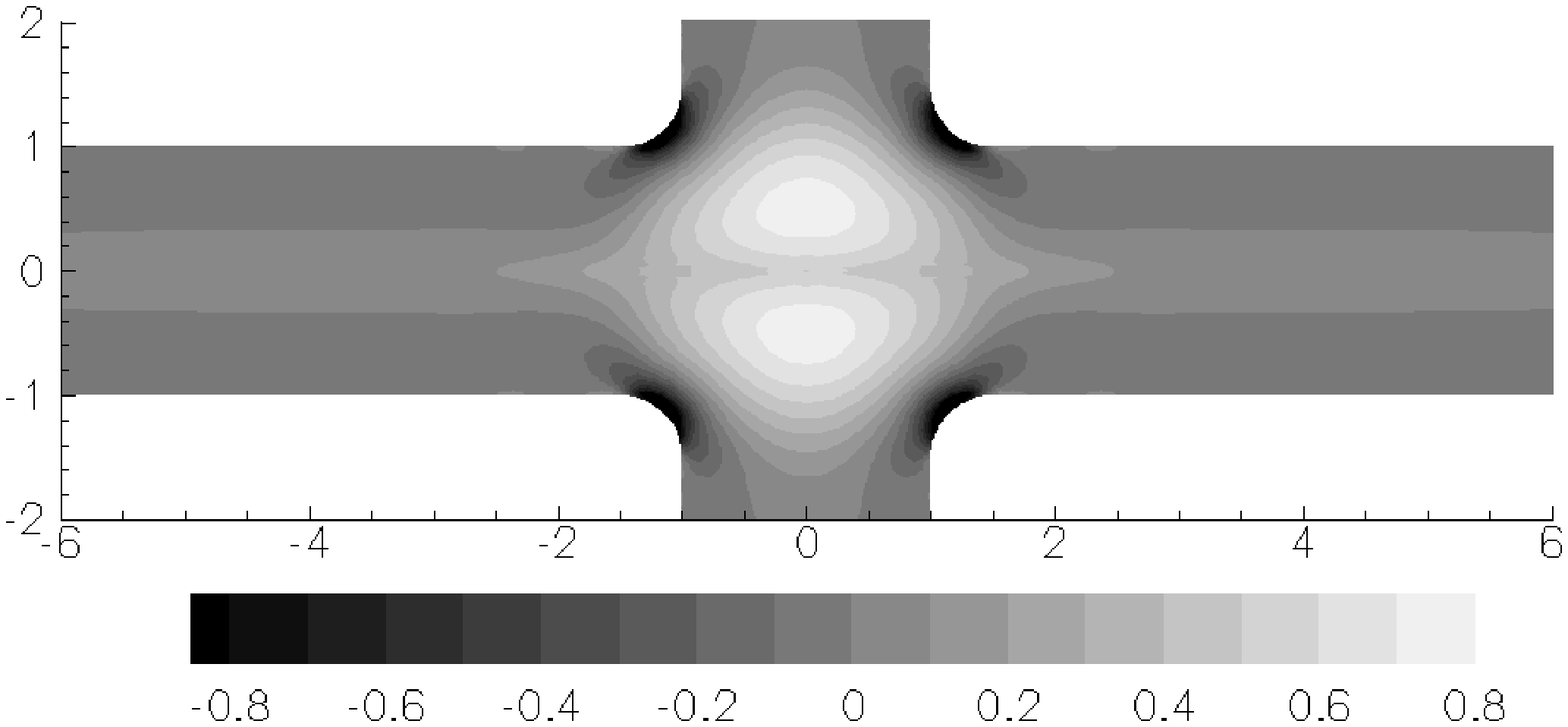}%
    }%
    \vspace{2pt}%
    \\%
    \subfigure[~$\textrm{tr}(\boldsymbol{\alpha})$]%
    {%
        \label{tra_ss_2006-10-12g}%
        \includegraphics[width=0.80\textwidth]{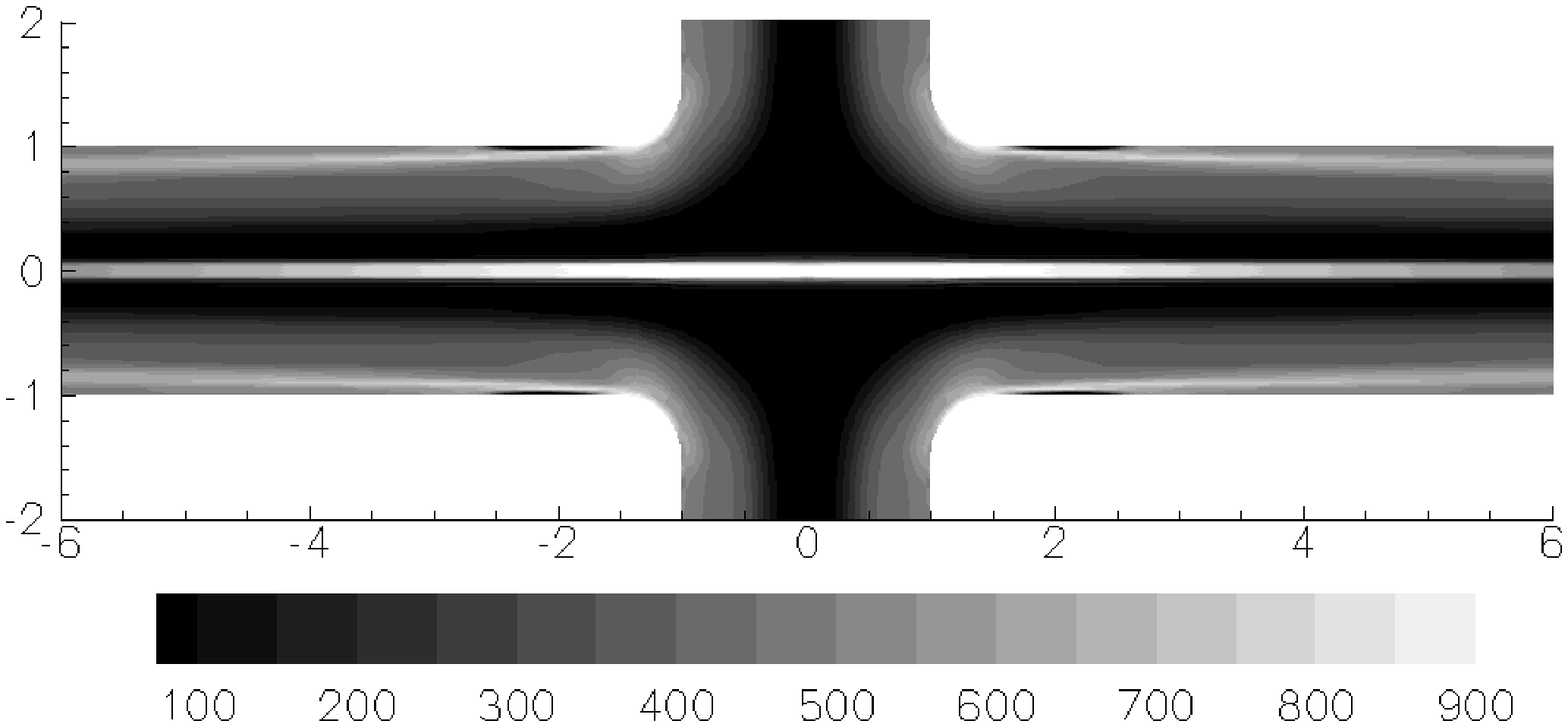}%
    }%
    \caption{Contour plots of steady state solution: $\textit{Wi} = 50$~(only the central part of the flow domain is shown).}%
    \label{ss_2006-10-12g}
\end{figure}

Steady state solutions are found for all $\textit{Wi}$ investigated
($0.2<\textit{Wi}<100$) in our study. For $\textit{Wi}\leqslant 60$
steady states are found by time integration and for those with
larger $\textit{Wi}$ Newton iteration (parameter continuation) is
used because of possible loss of stability, as we describe below. At
low $\textit{Wi}$ the velocity field is virtually unaffected by the
polymer molecules. Velocity contours at $\textit{Wi} = 0.2$ are
plotted in Figure~\ref{vel_ss_2006-10-12a}; for clarity only part of
the channel is shown. A stagnation point is found at the center of
the domain~($(0,0)$)
In both incoming and outgoing channels, the flow is almost the same
as pressure driven flow in a straight channel. No distinct
difference can be observed for the incoming and outgoing directions
in velocity field. Figure~\ref{u1x_ss_2006-10-12a} shows contours of
extension rate at $\textit{Wi} = 0.2$, in which a region dominated
by extensional flow is found near stagnation point. High extension
rate is also found near the corners due to the no-slip walls. The
magnitude of polymer stretching can be measured by the trace of its
conformation tensor $\textrm{tr}(\boldsymbol{\alpha})$, and is
plotted in Figure~\ref{tra_ss_2006-10-12a}. At low $\textit{Wi}$,
the extent to which polymers are deformed is barely noticeable,
but it can be clearly seen that polymers are primarily stretched in
either the extensional flow near the stagnation point and corners or
the shear flows near the walls. At high $\textit{Wi}$ ($\textit{Wi}
= 50$, Figure~\ref{ss_2006-10-12g}), the situation is very
different. Polymers are strongly stretched by the extensional flow
near the stagnation point and this stretching effect by extensional
flow overwhelms that of the shear flow. A distinct band of highly
stretched polymers (the birefringent strand)
forms~(Figure~\ref{tra_ss_2006-10-12g}). Since the polymer
relaxation time in this case is larger than the flow convection time
from stagnation point to the exits, this birefringent strand extends
 the whole length of the simulation domain. The resulting high polymer
stress significantly affects the velocity
field~(Figure~\ref{vel_ss_2006-10-12g}). Regions with reduced
velocity extend much farther away in the downstream directions of
the stagnation point than in the low $\textit{Wi}$ case, especially
along the x-axis, where high polymer stress dominates.
Correspondingly, a reduction in the extension rate near the
stagnation point is observed, most noticeably along the birefringent
strand (Figure~\ref{u1x_ss_2006-10-12g}).

\begin{figure}
    \centering%
    \includegraphics[width=0.75\textwidth]{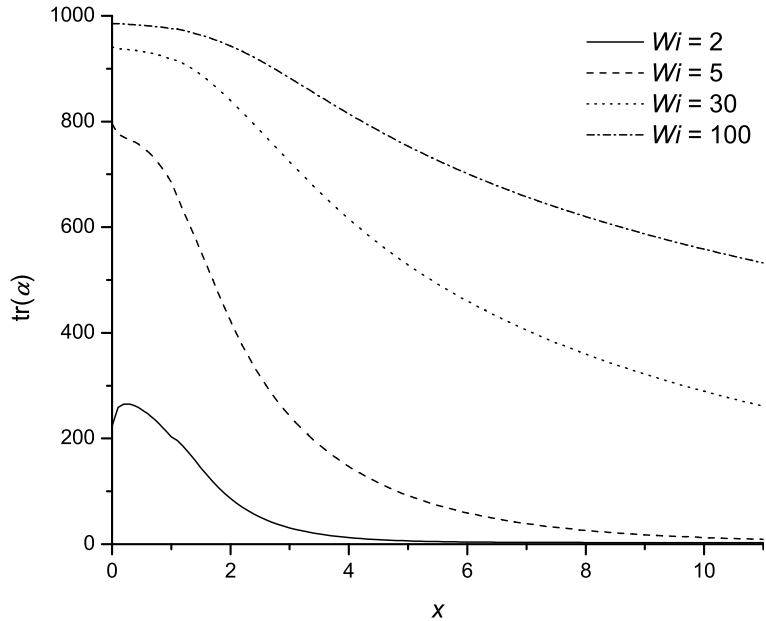}%
    \caption{Profile of $\textrm{tr}(\boldsymbol{\alpha})$ along $y = 0$.}%
    \label{strand_co_axial}
\end{figure}

\begin{figure}
    \centering%
    \includegraphics[width=0.75\textwidth]{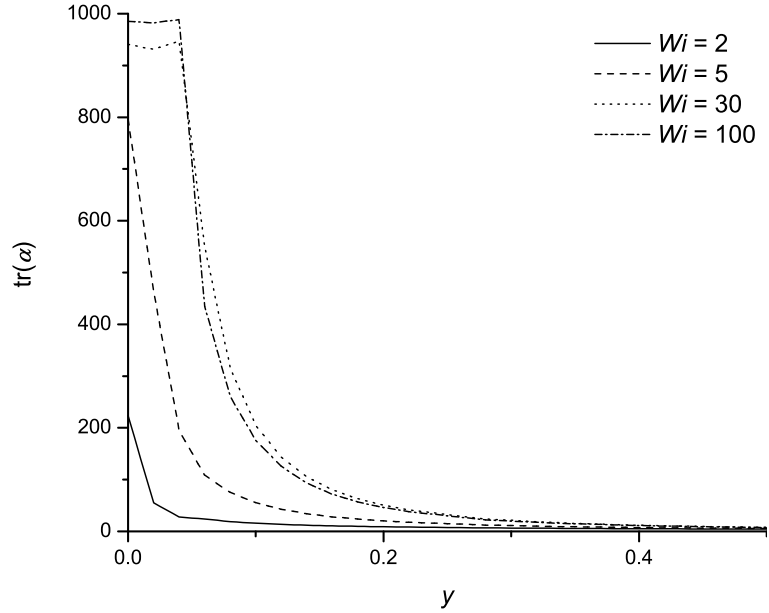}%
    \caption{Profile of $\textrm{tr}(\boldsymbol{\alpha})$ along $x = 0$ in the region very near the stagnation point.}%
    \label{strand_cross_sectional}
\end{figure}

\begin{figure}
    \centering%
    \includegraphics[width=0.75\textwidth]{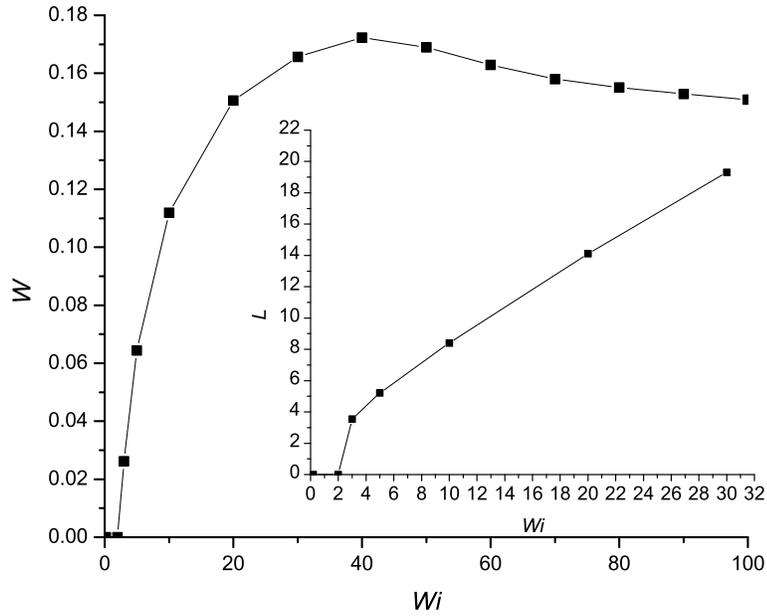}%
    \caption{Birefringence strand width $W$; Inset: Birefringence strand length $L$ ($\textrm{tr}(\boldsymbol{\alpha}) > 300$ is considered as observable birefringence region).}%
    \label{birefringence_width}
\end{figure}

Figures~\ref{strand_co_axial} and \ref{strand_cross_sectional} show
profiles at various values of $\textit{Wi}$ of
$\textrm{tr}(\boldsymbol{\alpha})$ along the outflow~(x-axis) and
inflow~(y-axis) directions of this stagnation point~(note the
difference in scales in the two plots). For increasing $\textit{Wi}$
the length of the region with highly stretched polymer keeps
increasing due to the increased relative relaxation
time~(Figure~\ref{strand_co_axial}). In high $\textit{Wi}$ cases
($\textit{Wi} = 30$ and $\textit{Wi} = 100$), polymers are not fully
relaxed even when they reach the exit of the simulation domain. The
cross-sectional view of $\textrm{tr}(\boldsymbol{\alpha})$ profile
along the y-axis~(Figure~\ref{strand_cross_sectional}) shows
interesting non-monotonic behaviors. Although the height of the
profile~($\textrm{tr}(\boldsymbol{\alpha})_{\textrm{max}}$) keeps
increasing upon increasing $\textit{Wi}$, the width of $\textit{Wi}
= 100$ case is smaller than that of $\textit{Wi} = 30$, resulting in
a steeper transition section between low and high stretching
regions. If we arbitrarily define
$\textrm{tr}(\boldsymbol{\alpha})>300$ as the observable
birefringence region, the width $W$ and the length $L$ of the
birefringent strand (measured on the inflow and outflow axes,
respectively) can be plotted as functions of $\textit{Wi}$, as in
Figure~\ref{birefringence_width}~(values of $L$ for $\textit{Wi}>30$
are not shown since they exceed the length of the simulation
domain). A clear non-monotonic trend is observed in the plot of
birefringence width, where $W$ increases sharply at relatively low
$\textit{Wi}$ and peaks around $\textit{Wi} = 40$. After that $W$
decreases mildly but consistently with further higher $\textit{Wi}$.
This non-monotonic trend is consistent with experimental
observations of birefringence in opposed-jet
devices~\citep{Muller_JNNFM1988}.

\begin{figure}
    \centering%
    \includegraphics[width=0.75\textwidth]{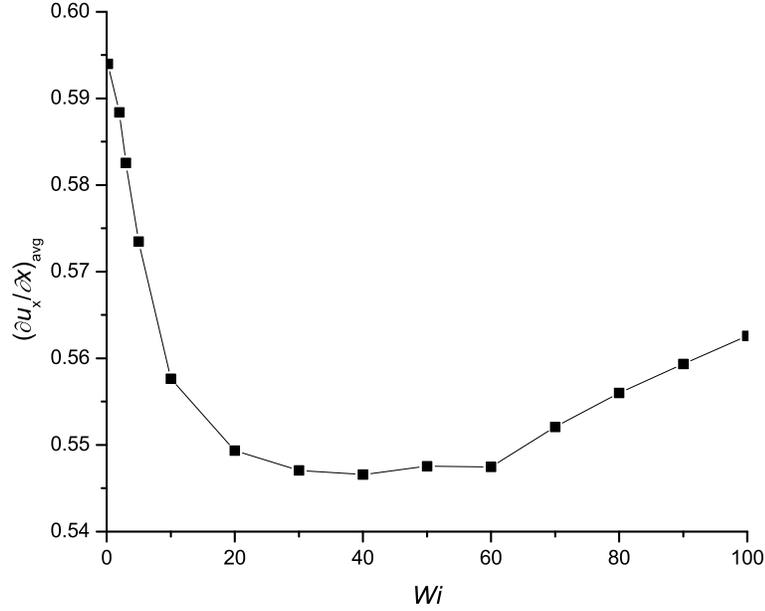}%
    \caption{Average extension rate $(\partial u_{x}/\partial x)_{avg}$ (averages taken in the domain $-0.1<x<0.1, -0.1<y<0.1$).}%
    \label{extension_rate}
\end{figure}

Similarly, a non-monotonicity is also found in the change of
velocity field with $\textit{Wi}$. Shown in
Figure~\ref{extension_rate} is the value of extension rate, averaged
within a box around the stagnation point ($-0.1<x<0.1, -0.1<y<0.1$),
as a function of $\textit{Wi}$. As $\textit{Wi}$ increases, the
extension rate decreases at low $\textit{Wi}$ but increases at high
$\textit{Wi}$, with a minimum found around $\textit{Wi}=40$.
Besides, most of experimental results are presented in terms of
Deborah number~($\textit{De}$), defined as the product of the
polymer relaxation time and an estimate of the extension rate near
the stagnation point. Noticing that the average~(nondimensionlized)
extension rate changes within a very narrow range~(around $0.55\sim
0.6$), a conversion $\textit{De}=0.3\textit{Wi}$ can be adopted for
comparison of our results with experimental ones.

Some understanding of this non-monotonicity can be gained by looking
at Figure~\ref{strand_co_axial}. Here it can be seen that for
$\textit{Wi}\lesssim 30$, the birefringent strand is not yet ``fully
developed'' in the sense that the polymer stretching is not yet
saturating near full extension. Thus the evolution of the velocity
field in this regime of $\textit{Wi}$ reflects the significant
changes that occur in the stress field in this regime. At higher
$\textit{Wi}$, however, the polymer stress field in the strand is
saturating, and thus not changing significantly. Furthermore, at
these high Weissenberg numbers, the relaxation of stress downstream
of the stagnation point diminishes, decreasing the gradient
$\partial\boldsymbol{\tau}_{xx}/\partial x$ and thus decreasing the
effect of viscoelasticity on the flow near the stagnation point.



\subsection{Periodic Orbits}
\begin{figure}%
    \centering%
    \includegraphics[width = 0.75\textwidth]{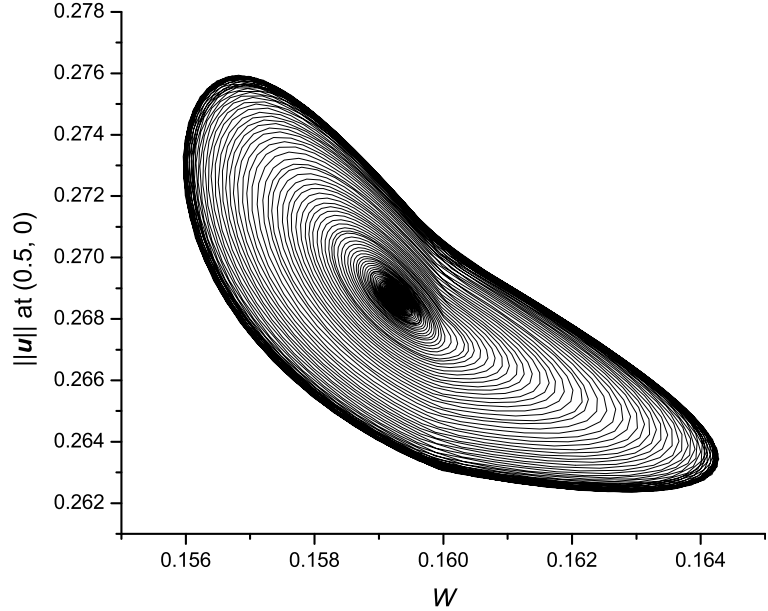}%
    \caption{Two dimensional projection of the dynamic trajectory from the steady state to the periodic orbit at $\textit{Wi}=66$: $\lVert \boldsymbol{u}\rVert =\sqrt{u_{x}^{2}+u_{y}^{2}}$ at $(0.5,0)$ v.s. $W$.}
    \label{velx0.5y0_birefringence_2006-11-24a}%
\end{figure}

We turn now to the stability of the steady states that have just
been described. Rather than attempting to compute the eigenspectra
of the linearization of the problem, an exceedingly demanding task,
we examine stability by direct time integration of perturbed steady
states. The perturbations take the form of slightly asymmetric
pressure profiles at the two entrances ($0.1\%$ maximum deviation
from the steady state value) that are applied for one time unit,
then released.
%
As an example, Figure~\ref{velx0.5y0_birefringence_2006-11-24a}
shows a two dimensional projection of the trajectory of the system
evolution over time at $\textit{Wi}=66$. Here the velocity magnitude
at a point near the stagnation point ($(0.5,0)$) is plotted against
the birefringent strand width $W$ measured on the inflow axis. The
system starts at the steady state with $W = 0.1593$ and
$\lVert\boldsymbol{u}\rVert _{(0.5,0)}= 0.2687$ and spirals outward
with time after the perturbation. Eventually the trajectory merges
into a cycle (the outer dark cycle in the
Figure~\ref{velx0.5y0_birefringence_2006-11-24a}). This clearly
identifies the existence of a stable periodic orbit. Note the
anticorrelation between $\lVert \boldsymbol{u}\rVert$ and $W$, i.e.
when the flow speeds up near the stagnation point, the strand thins
and vice versa.


\begin{figure}
    \centering%
    \includegraphics[width = 0.8\textwidth]{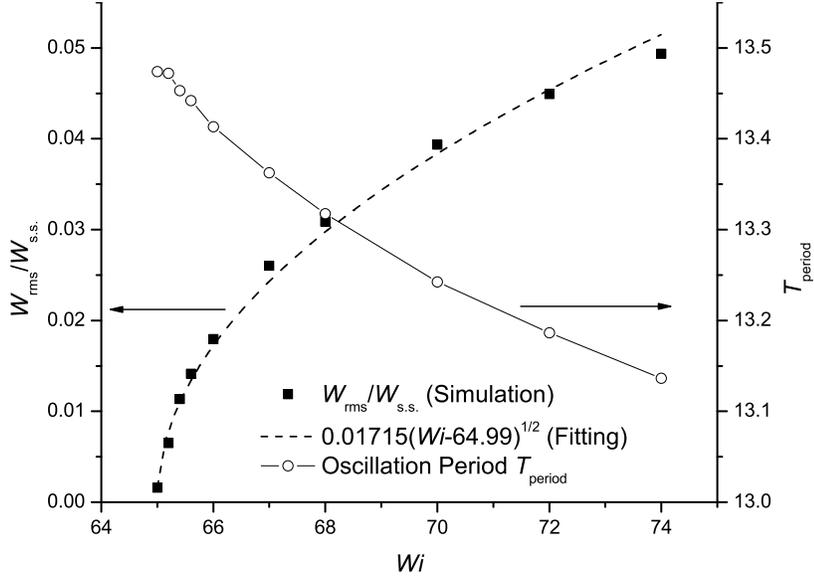}%
    \caption{Left: Root mean square deviations of the  birefringent strand width $W$ at periodic orbits, normalized by steady state values; Right: Oscillation periods.}%
    \label{birefringence_rms}%
\end{figure}

Figure~\ref{birefringence_rms} shows the root-mean-square deviations
over one period of $W$ from its steady state values
, normalized by the corresponding steady state values
$W_{\textrm{s.s.}}$, as a function of $\textit{Wi}$ for all the
cases where we found periodic orbits. Time integrations for
$\textit{Wi}>74$ did not converge due to the enormous stress
gradient around the corners of the no-slip walls and the consequent
numerical oscillations downstream. Data points for
$W_{\textrm{rms}}$ computed from our simulations are fitted with a
function of the form $a(\textit{Wi}-b)^{c}$, with $c$ fixed at
$1/2$. Very good agreement is found for our simulation data with the
$1/2$ power law, characteristic of a supercritical Hopf
bifurcation~\citep{Guckenheimer_1983}. The critical Weissenberg
number $\textit{Wi}_{\textrm{critical}}$ is identified to be $64.99$
by this fitting. Also shown in Figure~\ref{birefringence_rms} are
periods of oscillations, where a slight decrease with increasing
$\textit{Wi}$ is found. This is interesting since it indicates that
some time scale other than the polymer relaxation time sets the
period of oscillations.


Time-dependent fluctuations of birefringence width are also reported
in the experiments done by
M\"{u}ller~\textit{et~al.}~\citep{Muller_JNNFM1988} in opposed-jet
devices. In their optical experiments with semi-dilute aPS
solutions, the width of the birefringent strand oscillates rapidly
between two values in a certain range of extension rate. The
critical value of extension rate for the instability in their study
is close to the one where the birefringent strand width $W$ is
highest, while in our simulations $\textit{Wi}_{\textrm{critical}}$
($=65$) is somewhat larger than
the one ($\textit{Wi}\approx 40$) that gives the largest $W$.

\subsection{Instability Mechanism}
\begin{figure}
    \centering%
    \subfigure[$~t = 0$]%
    {%
        \label{ux_a}%
        \includegraphics[width = 0.29\textheight, bb = 22 253 587 575]{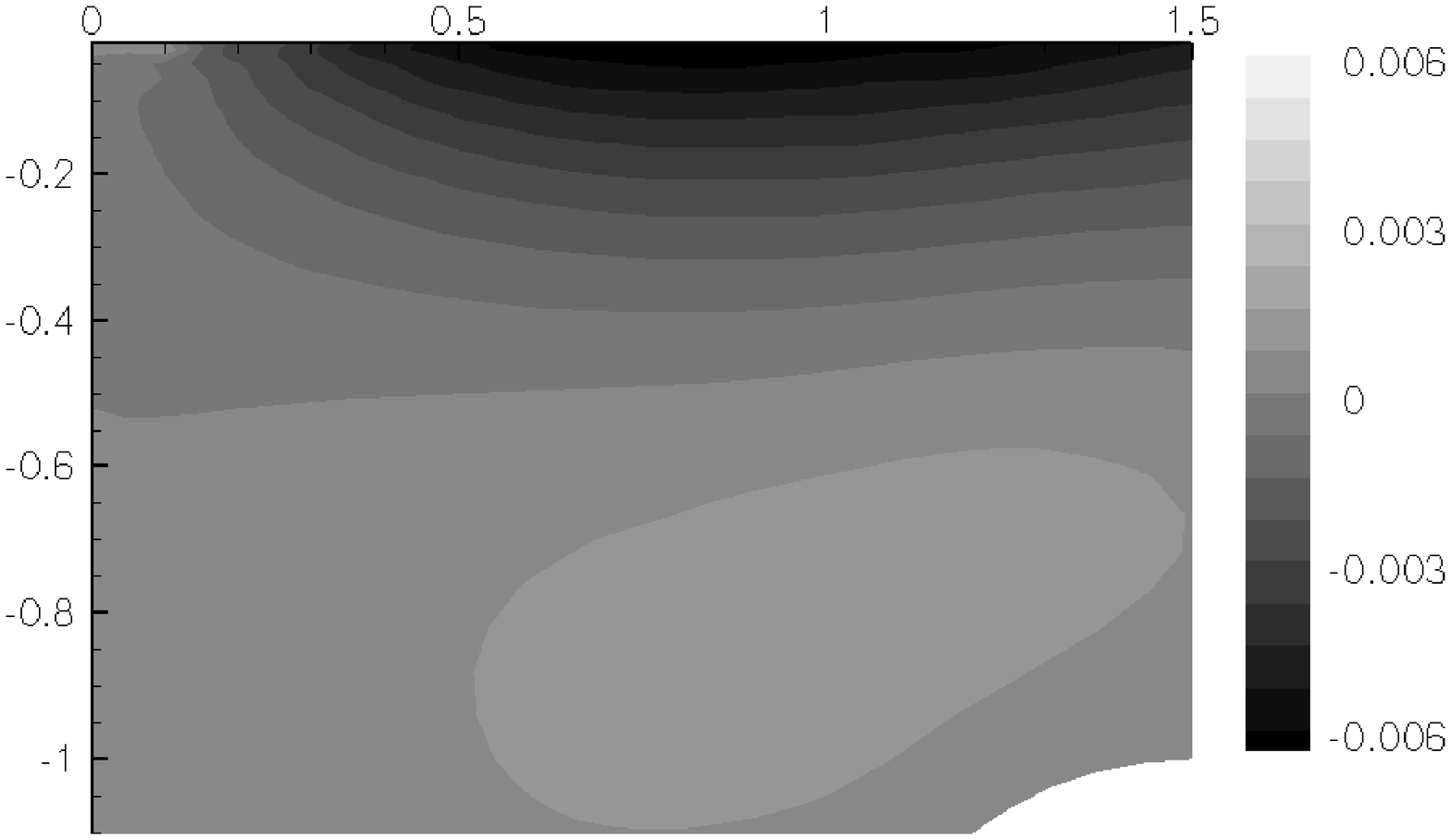}%
    }%
    \hspace{2pt}%
    \subfigure[$~t = 1.68$]%
    {%
        \label{ux_b}%
        \includegraphics[width = 0.29\textheight, bb = 22 253 587 575]{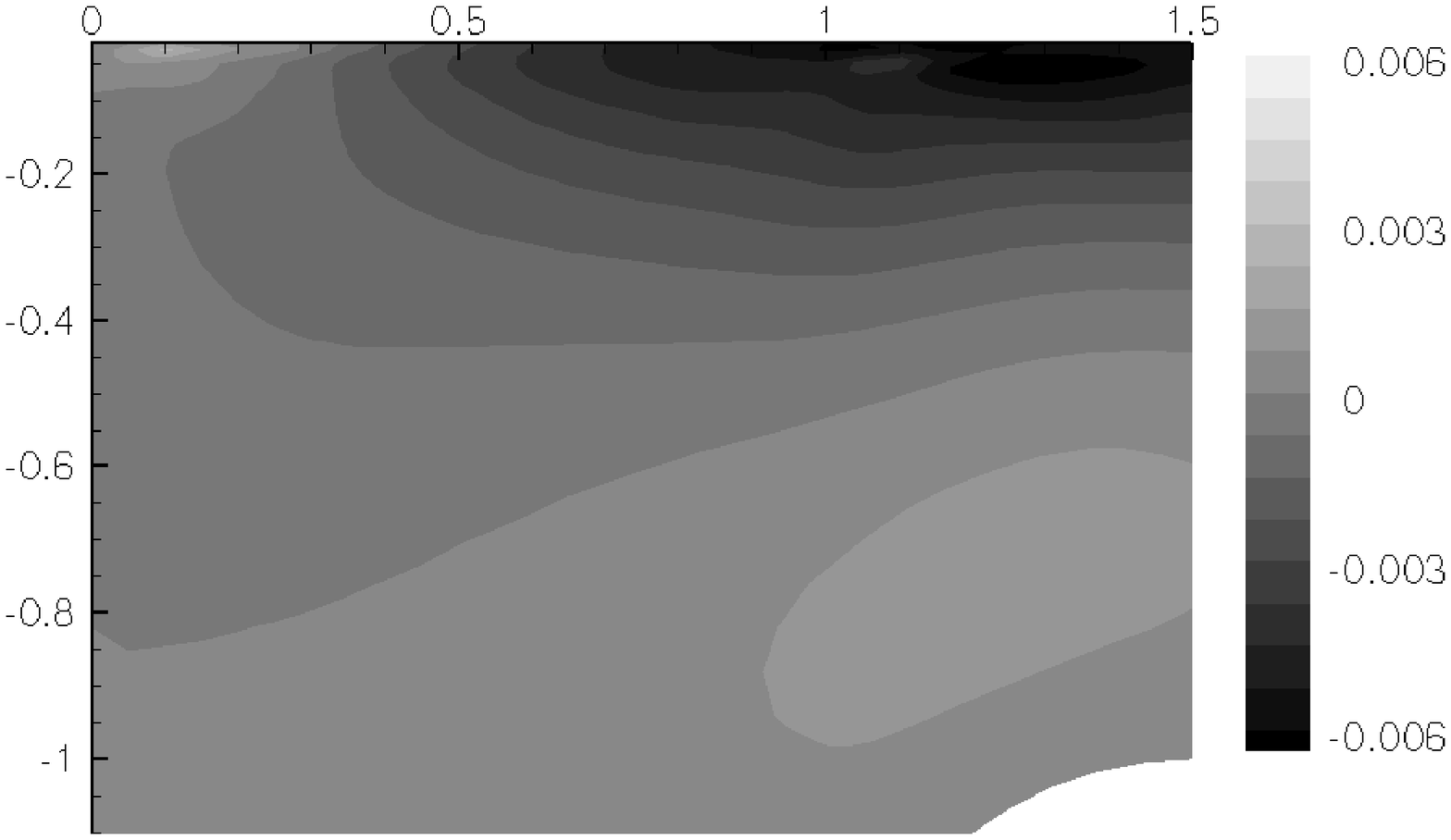}%
    }%
    \vspace{0pt}%
    \\%
    \subfigure[$~t = 3.35$]%
    {%
        \label{ux_c}%
        \includegraphics[width = 0.29\textheight, bb = 22 253 587 575]{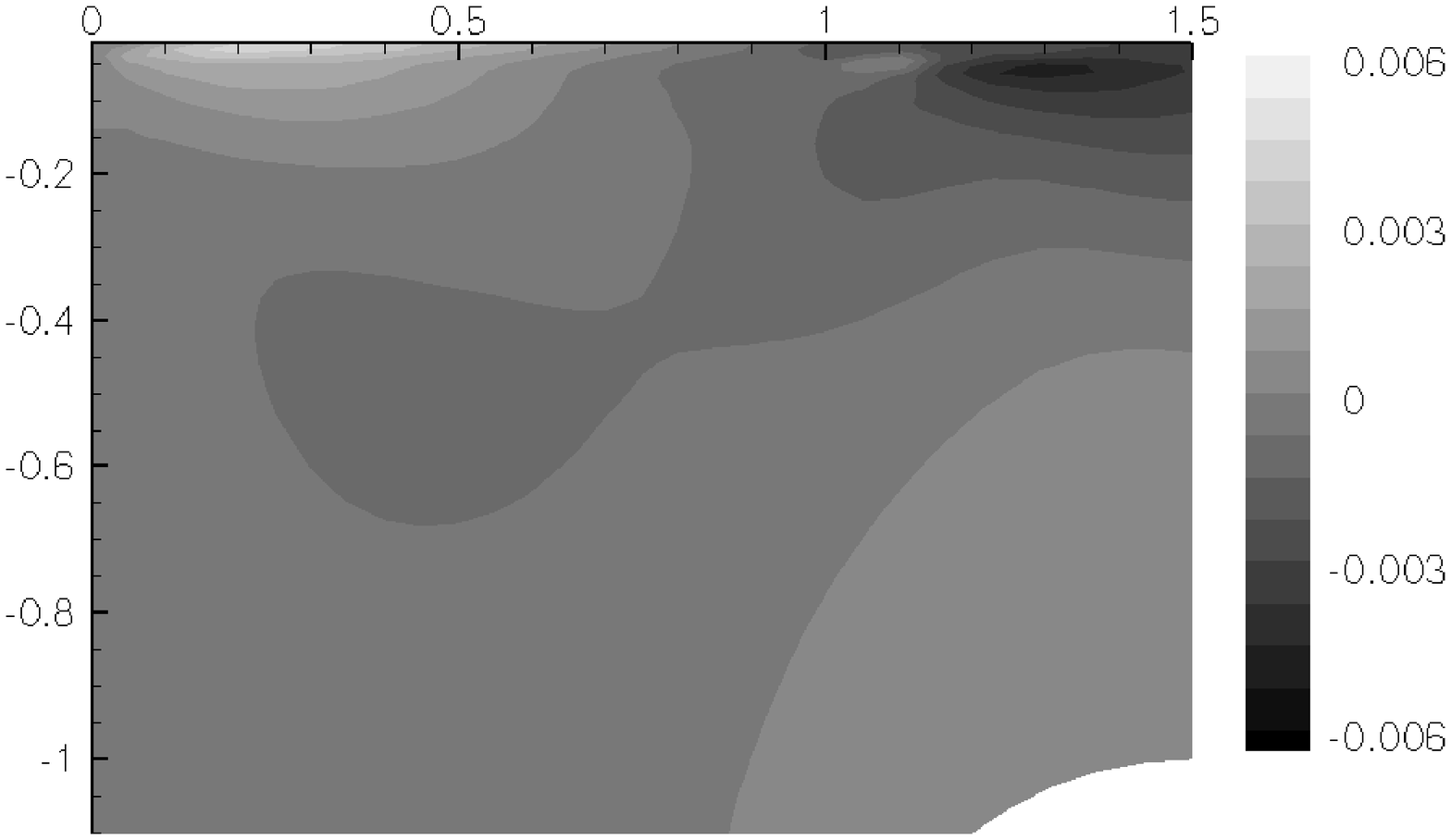}%
    }%
    \hspace{2pt}%
    \subfigure[$~t = 5.03$]%
    {%
        \label{ux_d}%
        \includegraphics[width = 0.29\textheight, bb = 22 253 587 575]{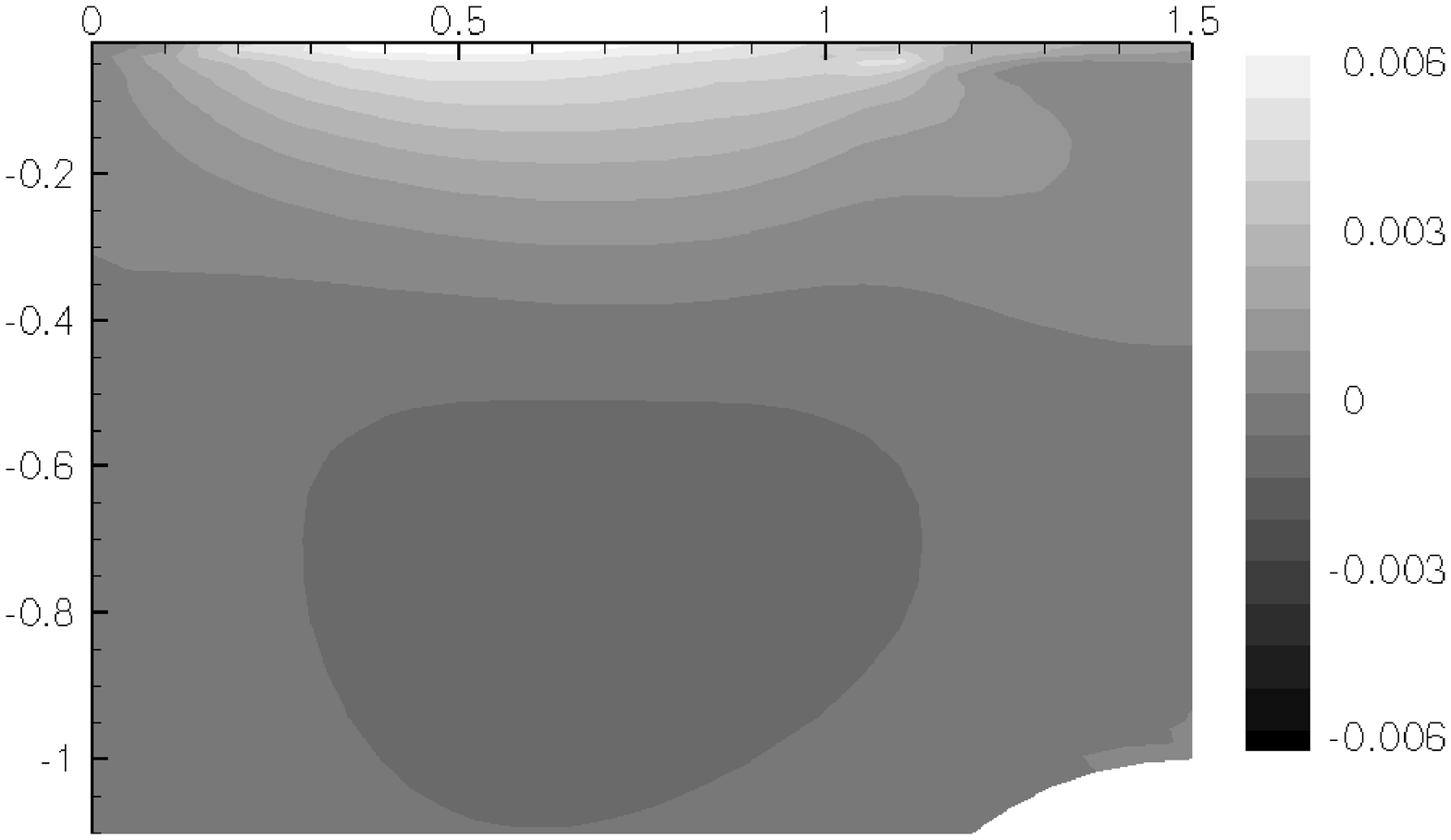}%
    }%
    \vspace{0pt}%
    \\%
    \subfigure[$~t = 6.71$]%
    {%
        \label{ux_e}%
        \includegraphics[width = 0.29\textheight, bb = 22 253 587 575]{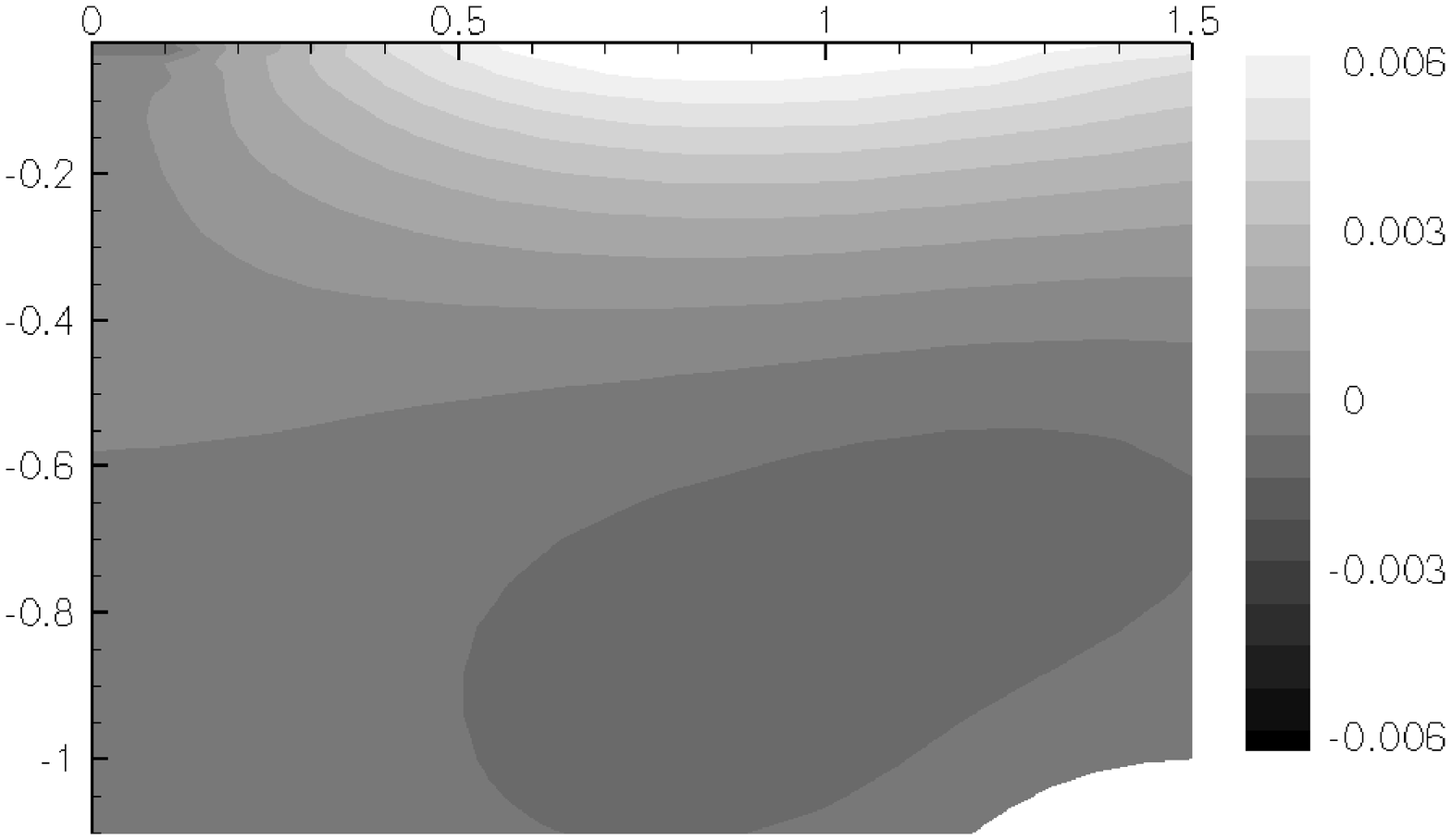}%
    }%
    \hspace{2pt}%
    \subfigure[$~t = 8.38$]%
    {%
        \label{ux_f}%
        \includegraphics[width = 0.29\textheight, bb = 22 253 587 575]{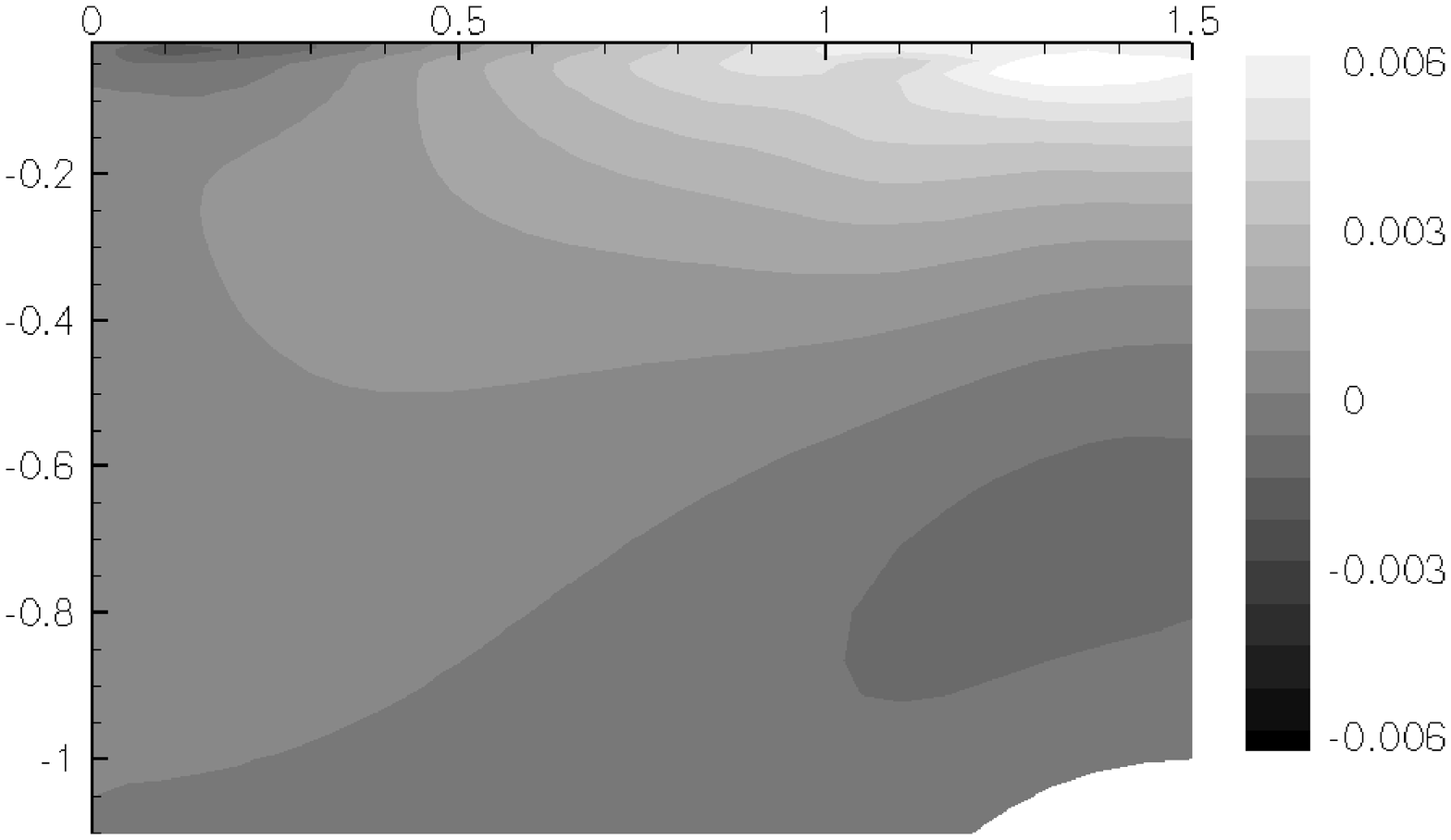}%
    }%
    \vspace{0pt}%
    \\%
    \subfigure[$~t = 10.06$]%
    {%
        \label{ux_g}%
        \includegraphics[width = 0.29\textheight, bb = 22 253 587 575]{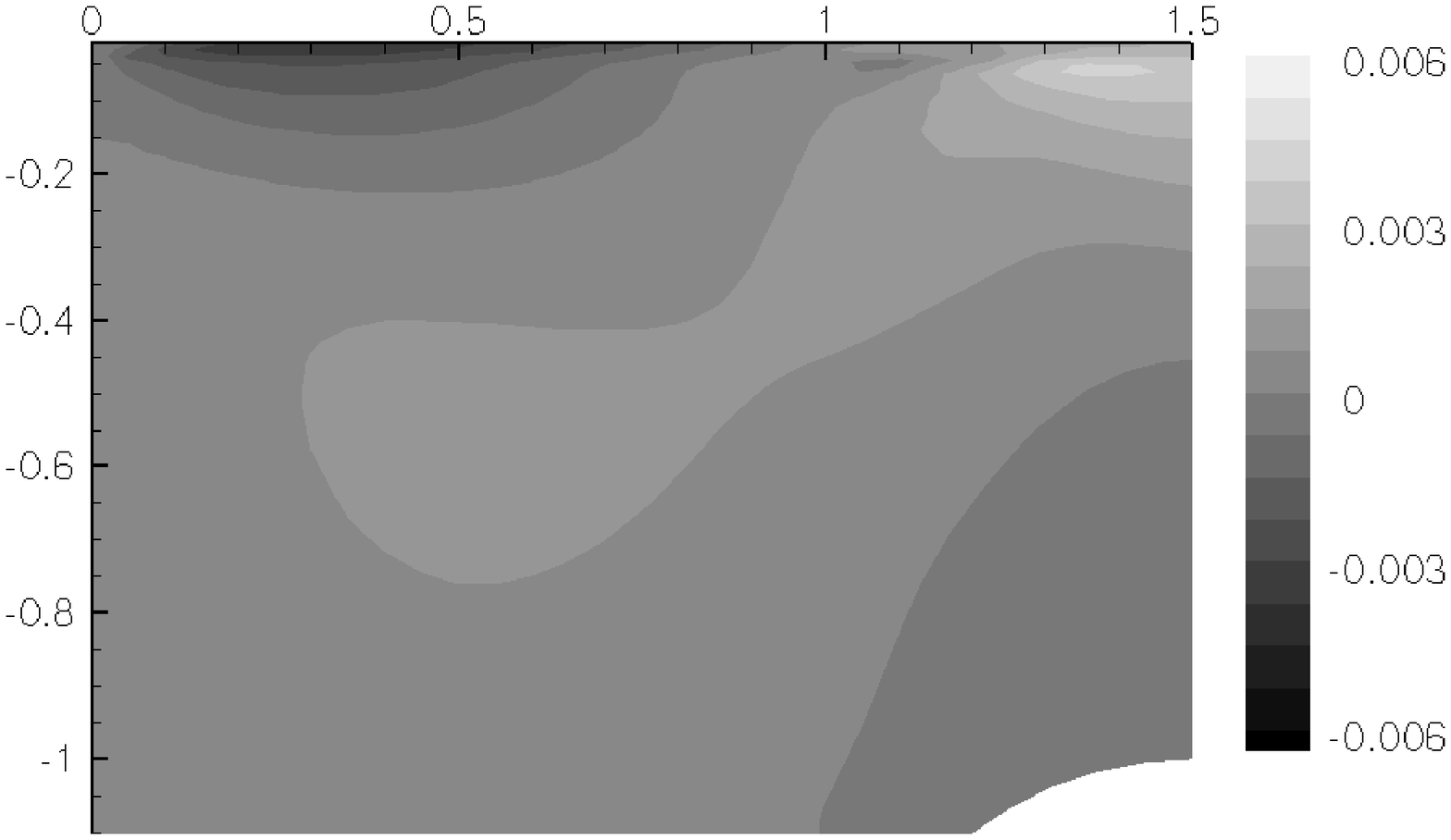}%
    }%
    \hspace{2pt}%
    \subfigure[$~t = 11.74$]%
    {%
        \label{ux_h}%
        \includegraphics[width = 0.29\textheight, bb = 22 253 587 575]{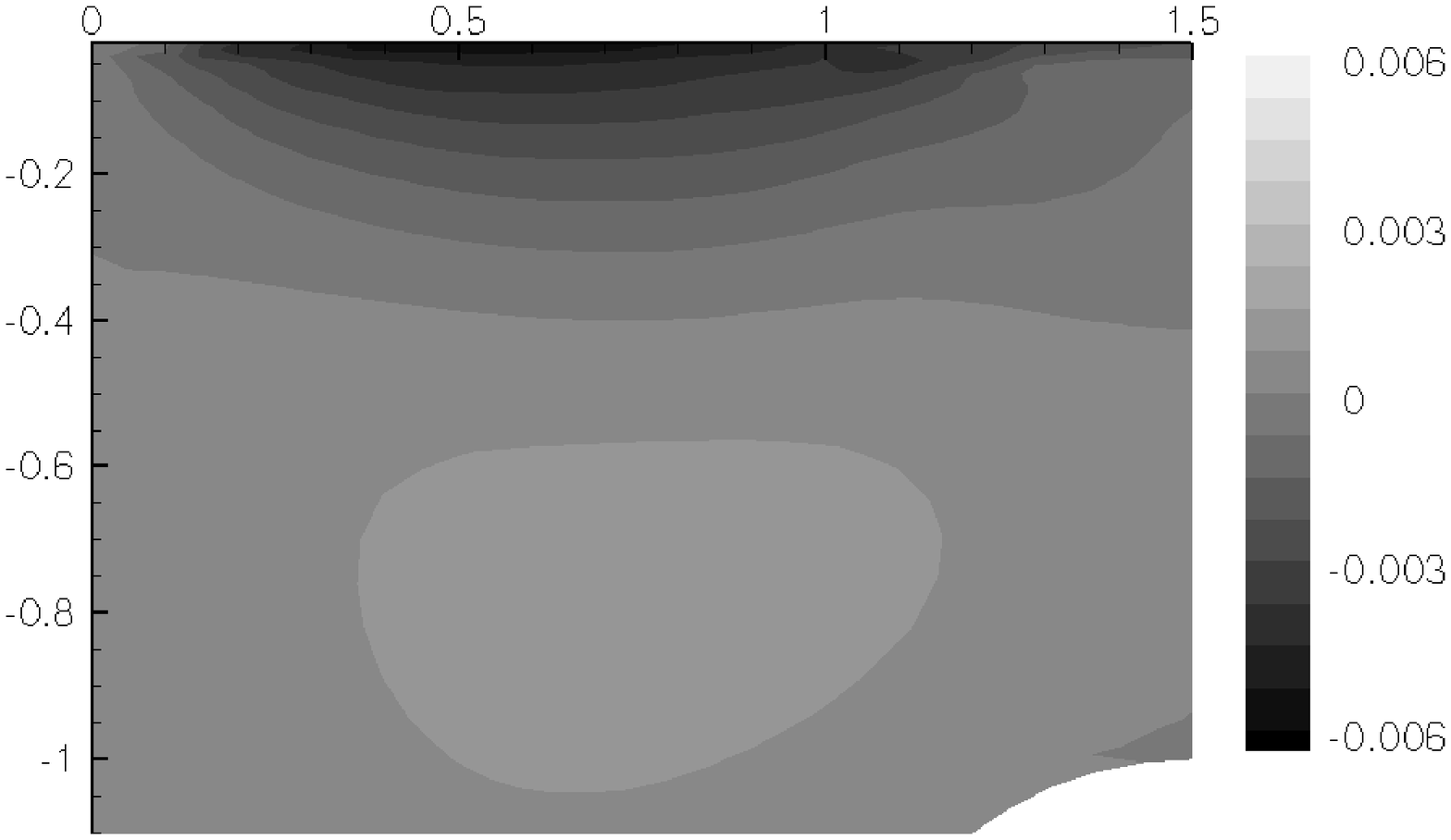}%
    }%
    \caption{Perturbation of x-component of velocity, $u_{x}'$ with respect to steady states at periodic orbits: $\textit{Wi} = 66$. The region shown is $0<x<1.5$, $-1.1<y<0$, stagnation point is at the top-left corner.}%
    \label{ux_2006-11-5a}
\end{figure}

\begin{figure}
    \centering%
    \subfigure[$~t = 0$]%
    {%
        \label{uy_a}%
        \includegraphics[width = 0.29\textheight, bb = 22 253 587 575]{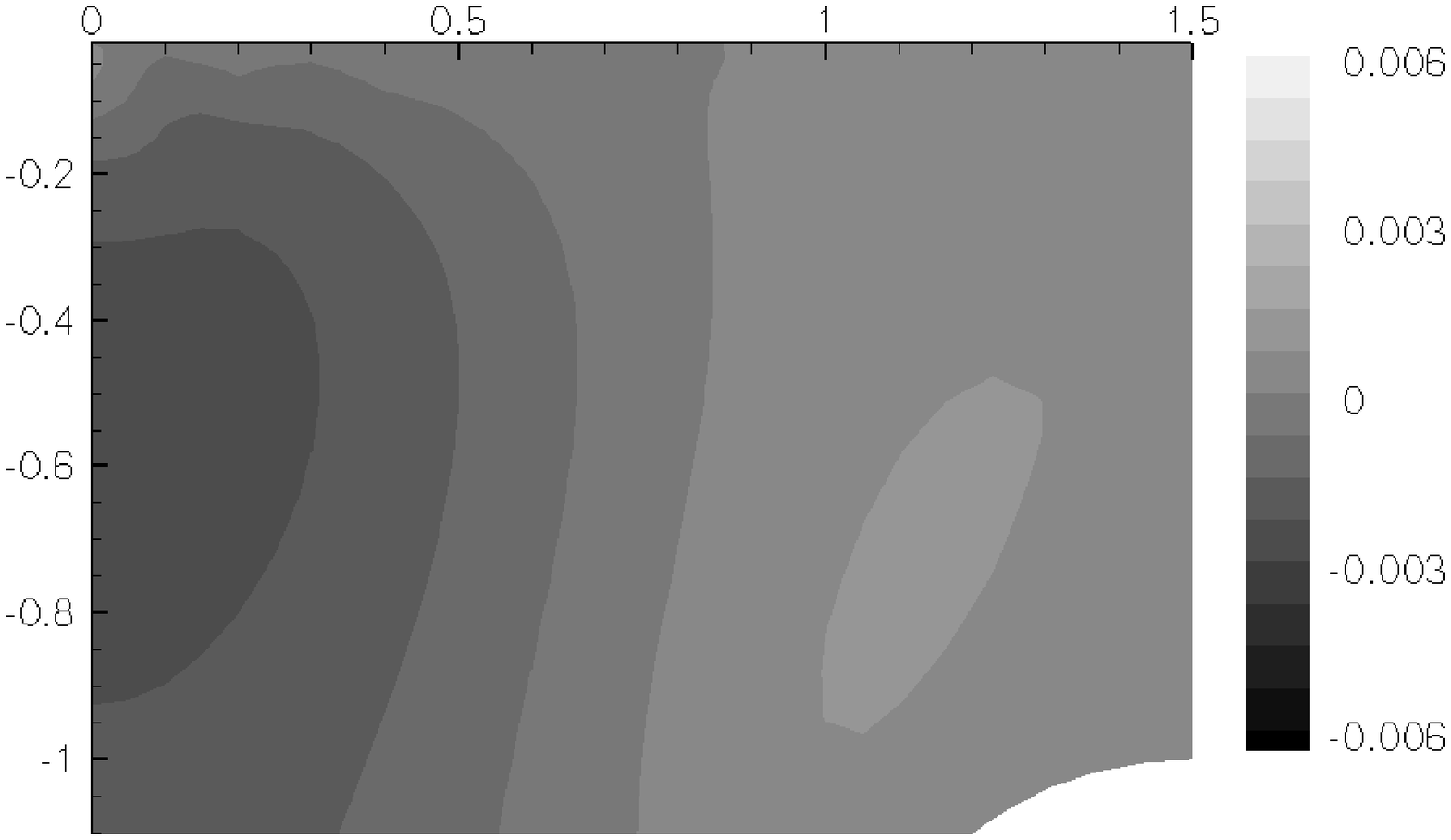}%
    }%
    \hspace{2pt}%
    \subfigure[$~t = 1.68$]%
    {%
        \label{uy_b}%
        \includegraphics[width = 0.29\textheight, bb = 22 253 587 575]{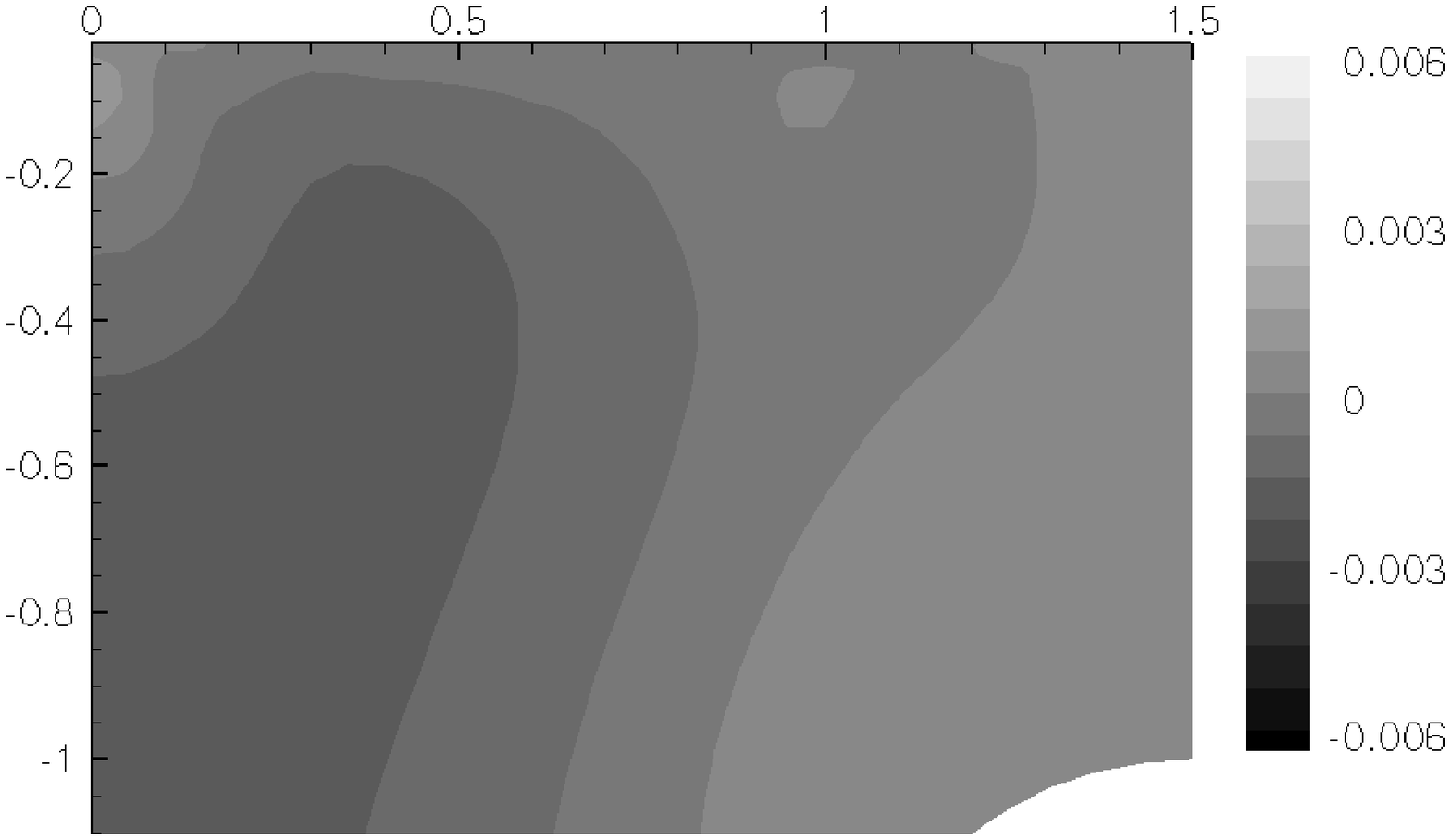}%
    }%
    \vspace{0pt}%
    \\%
    \subfigure[$~t = 3.35$]%
    {%
        \label{uy_c}%
        \includegraphics[width = 0.29\textheight, bb = 22 253 587 575]{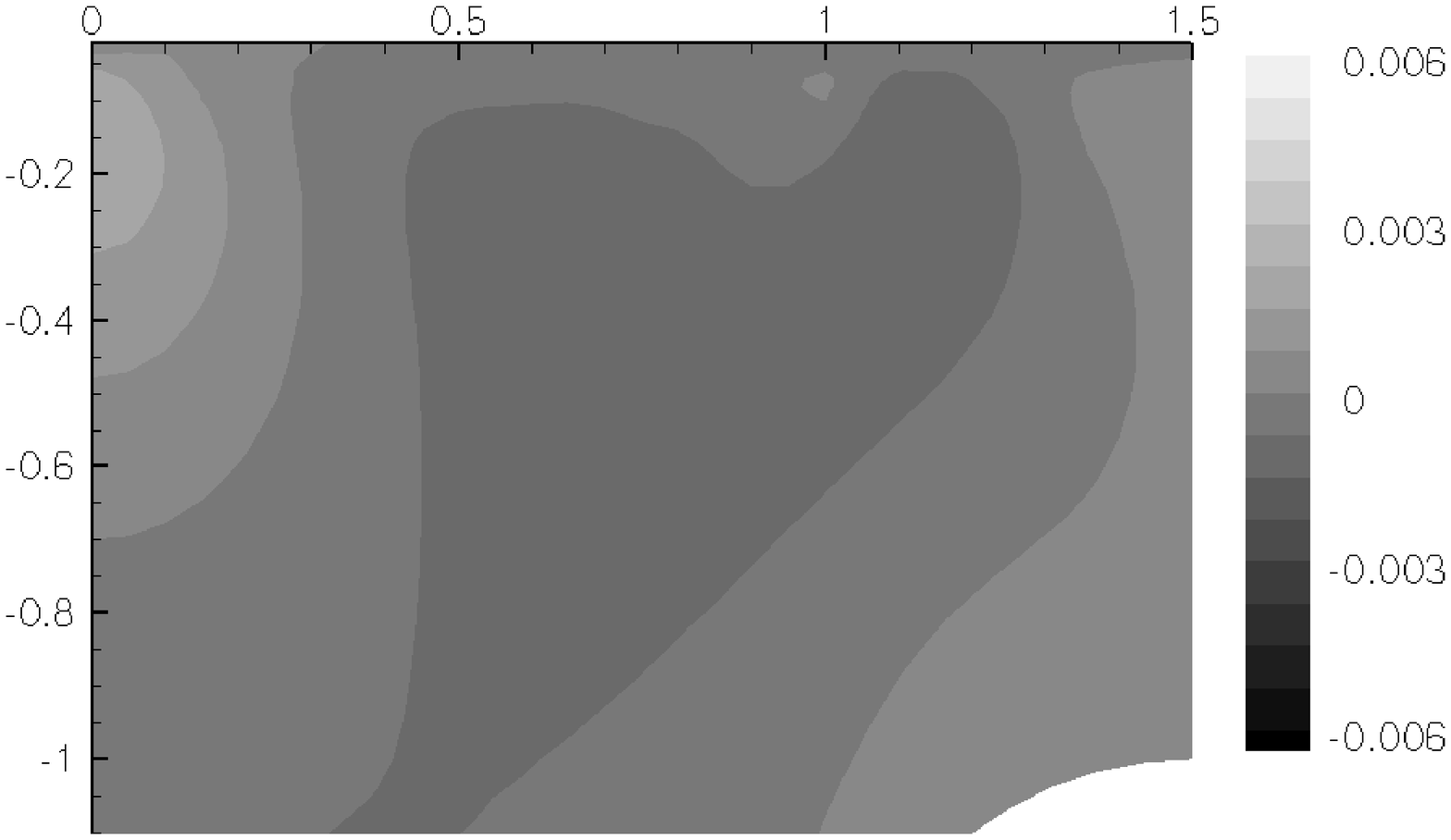}%
    }%
    \hspace{2pt}%
    \subfigure[$~t = 5.03$]%
    {%
        \label{uy_d}%
        \includegraphics[width = 0.29\textheight, bb = 22 253 587 575]{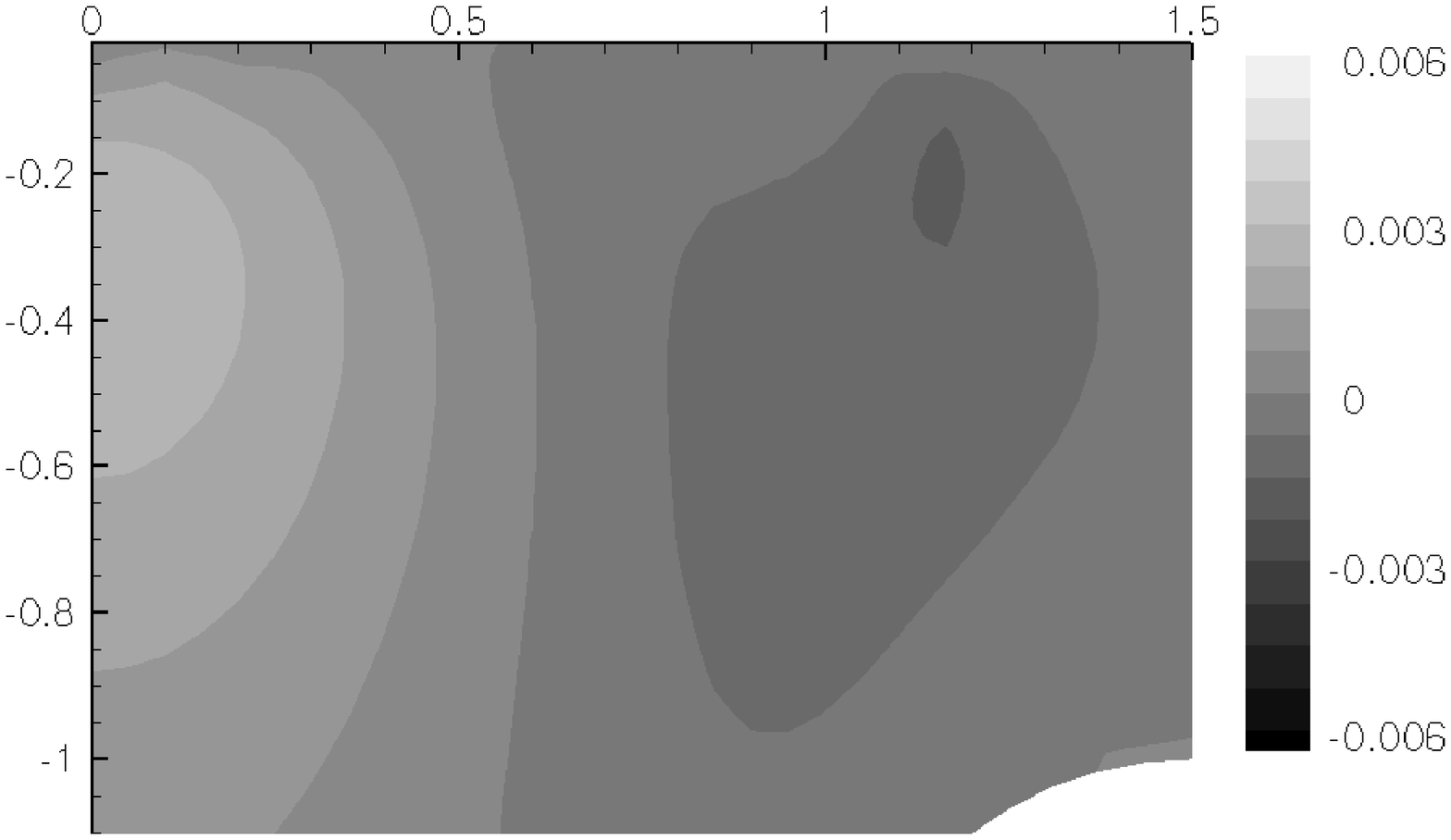}%
    }%
    \vspace{0pt}%
    \\%
    \subfigure[$~t = 6.71$]%
    {%
        \label{uy_e}%
        \includegraphics[width = 0.29\textheight, bb = 22 253 587 575]{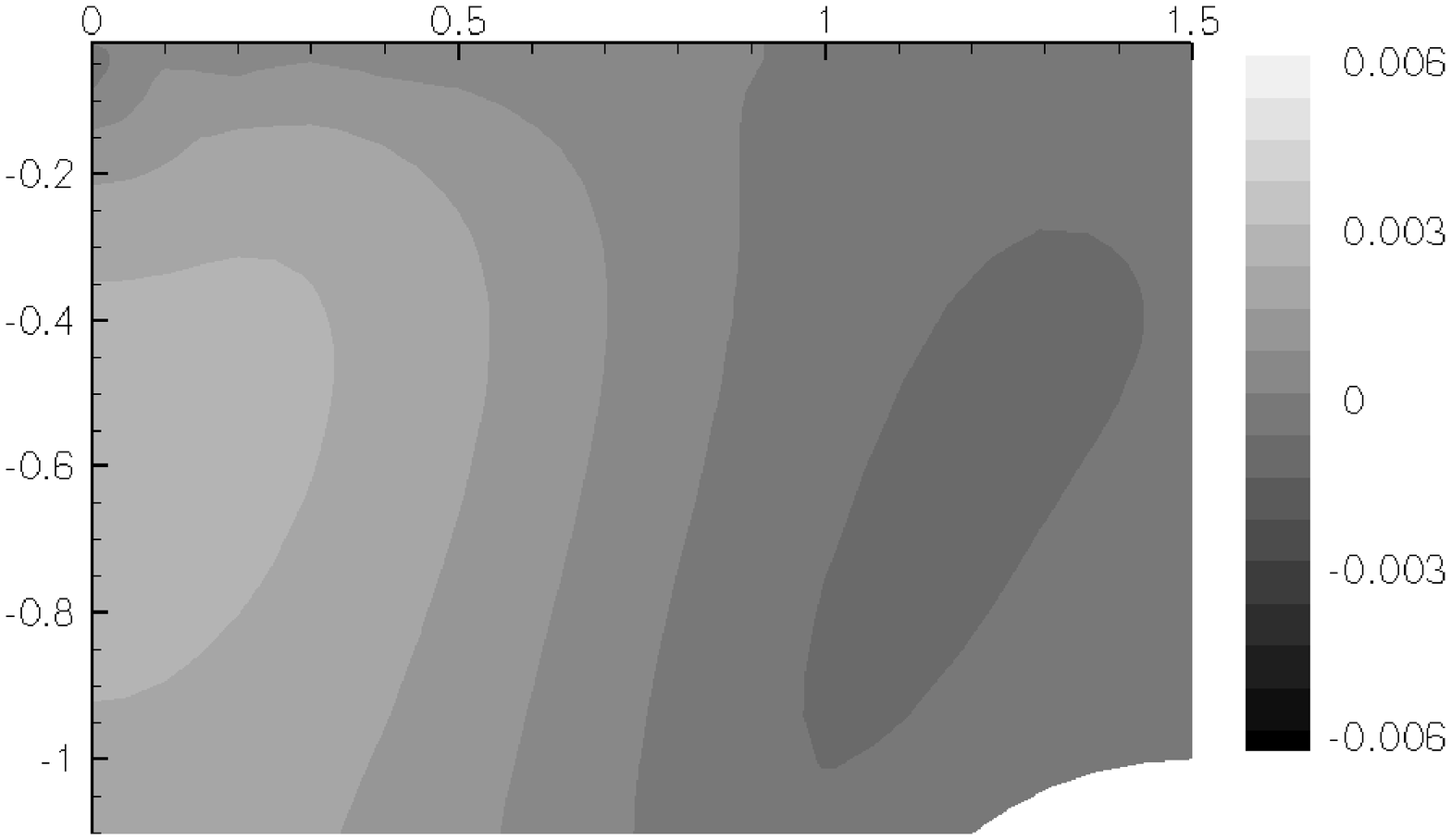}%
    }%
    \hspace{2pt}%
    \subfigure[$~t = 8.38$]%
    {%
        \label{uy_f}%
        \includegraphics[width = 0.29\textheight, bb = 22 253 587 575]{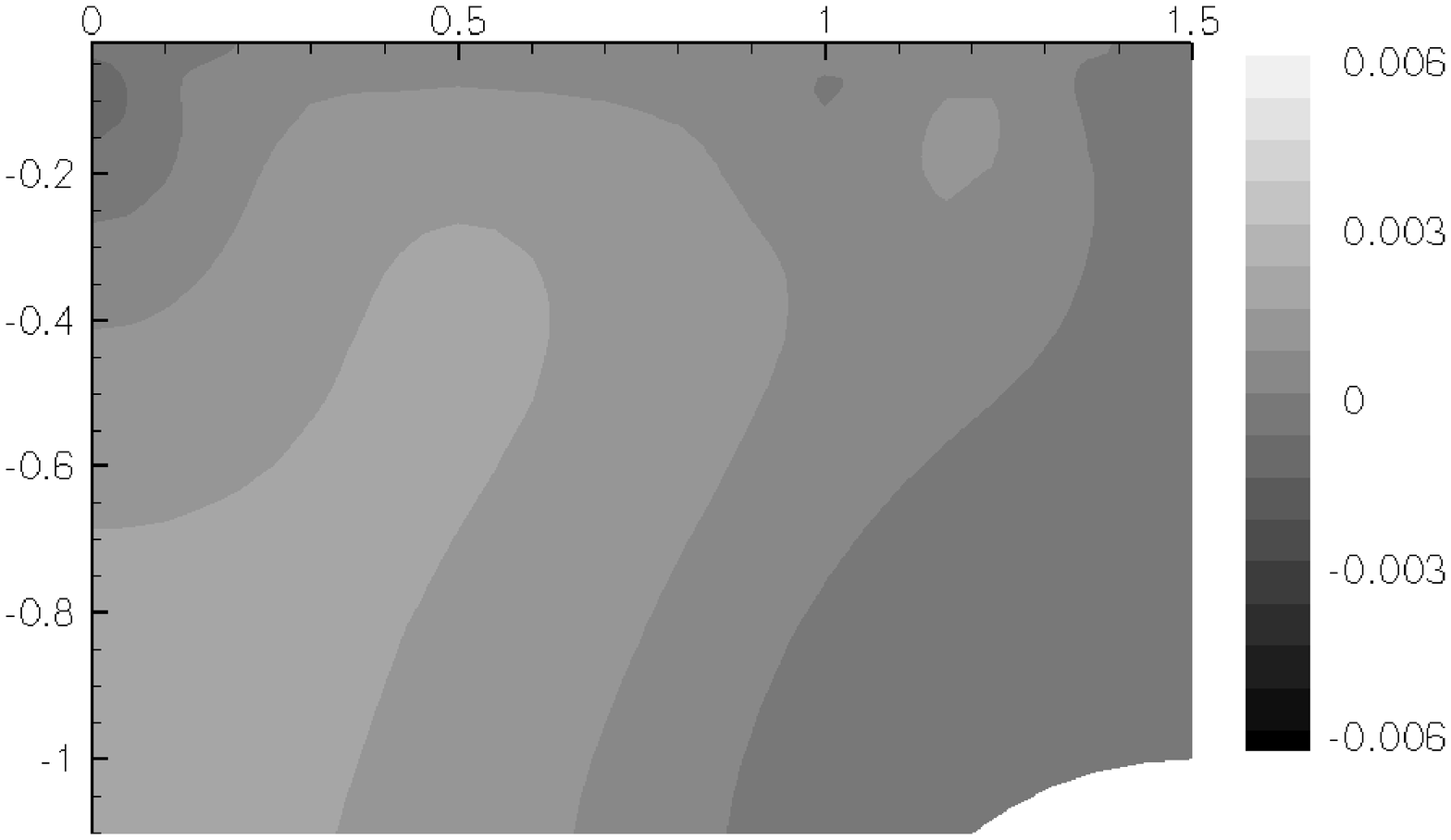}%
    }%
    \vspace{0pt}%
    \\%
    \subfigure[$~t = 10.06$]%
    {%
        \label{uy_g}%
        \includegraphics[width = 0.29\textheight, bb = 22 253 587 575]{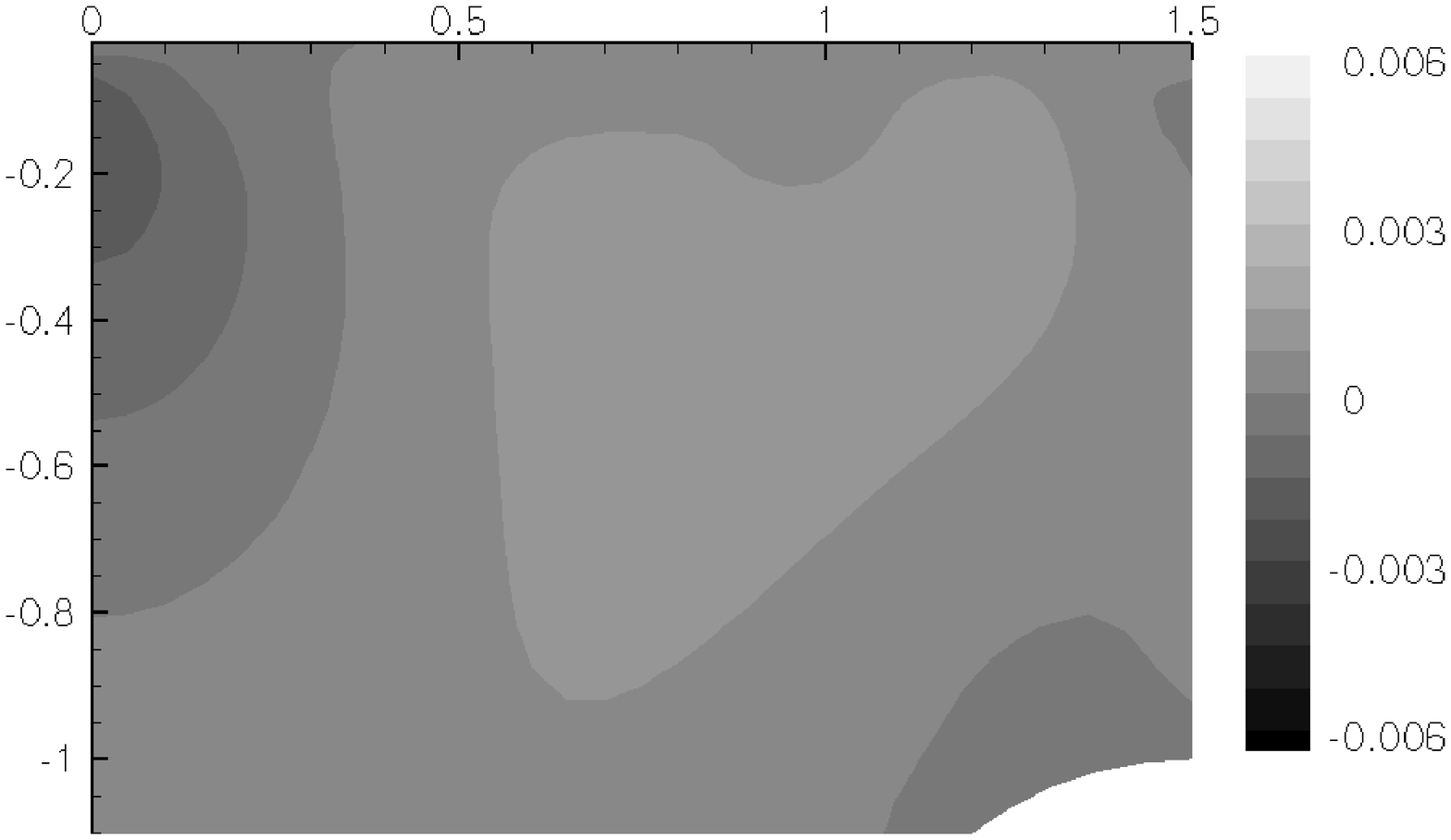}%
    }%
    \hspace{2pt}%
    \subfigure[$~t = 11.74$]%
    {%
        \label{uy_h}%
        \includegraphics[width = 0.29\textheight]{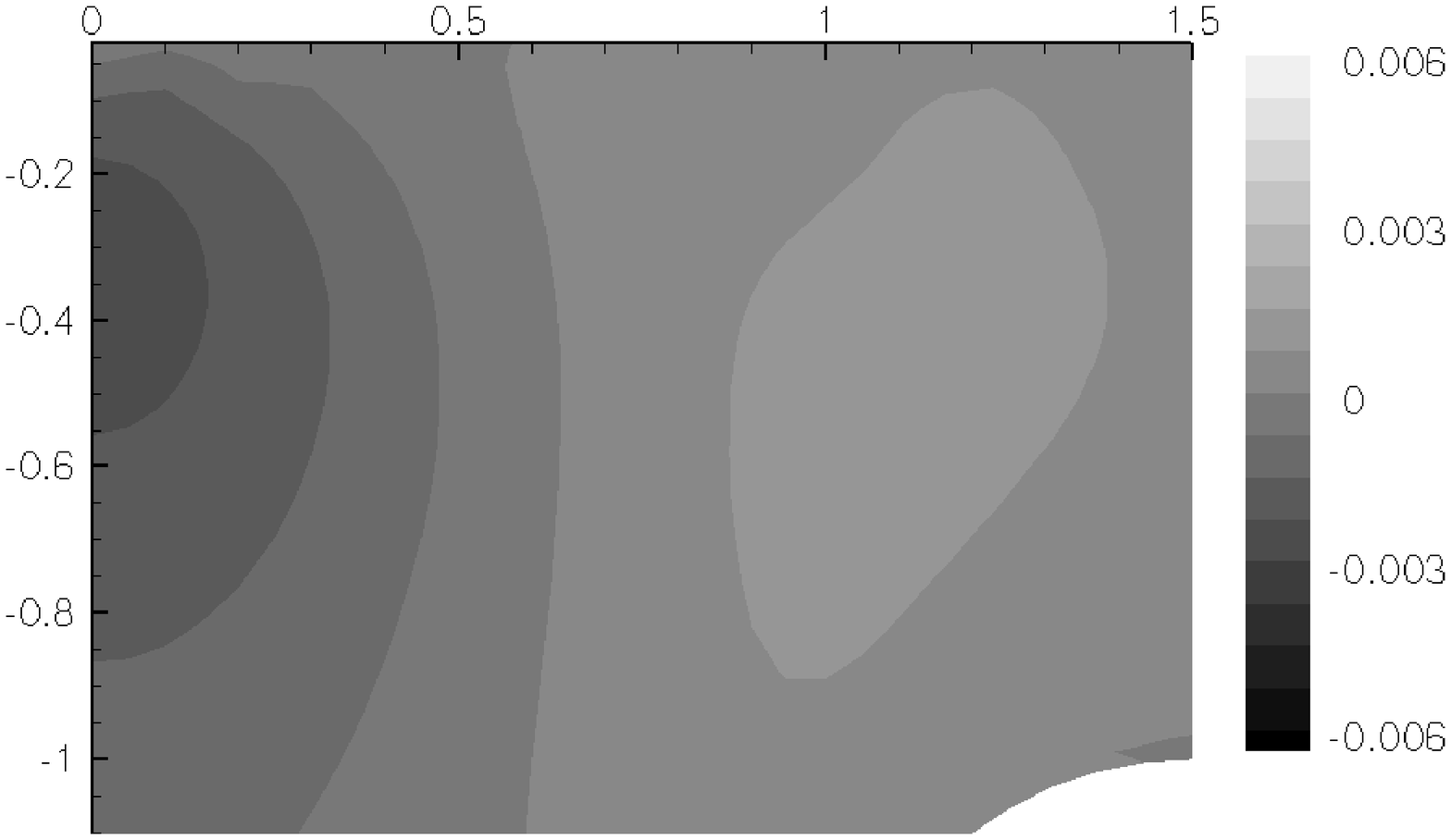}%
    }%
    \caption{Perturbation of y-component of velocity, $u_{y}'$ with respect to steady states at periodic orbits: $\textit{Wi} = 66$. The region shown is $0<x<1.5$, $-1.1<y<0$, stagnation point is at the top-left corner.}%
    \label{uy_2006-11-5a}
\end{figure}

\begin{figure}
    \centering%
    \subfigure[$~t = 0$]%
    {%
        \label{axx_a}%
        \includegraphics[width = 0.29\textheight, bb = 22 253 587 575]{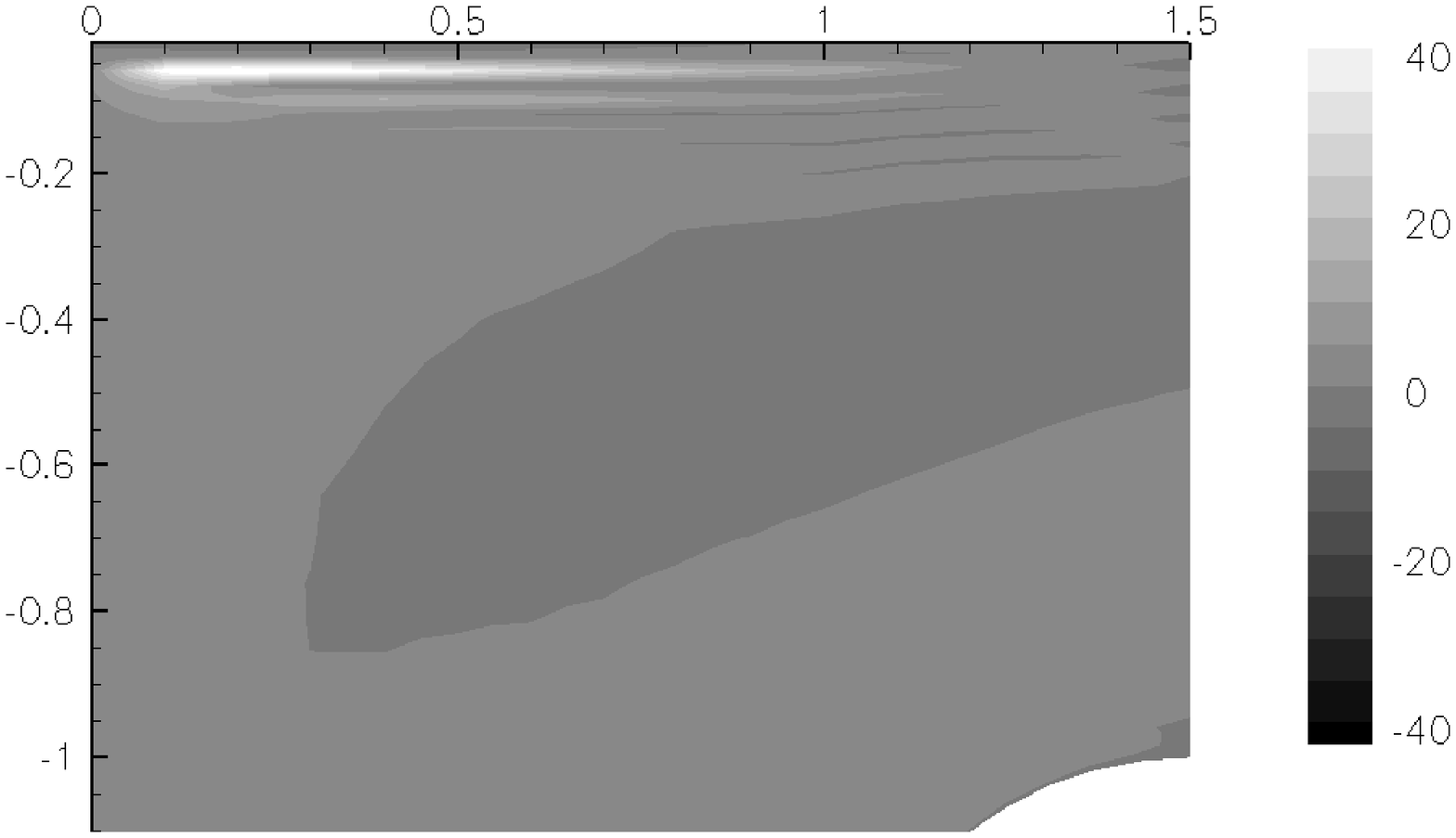}%
    }%
    \hspace{2pt}%
    \subfigure[$~t = 1.68$]%
    {%
        \label{axx_b}%
        \includegraphics[width = 0.29\textheight, bb = 22 253 587 575]{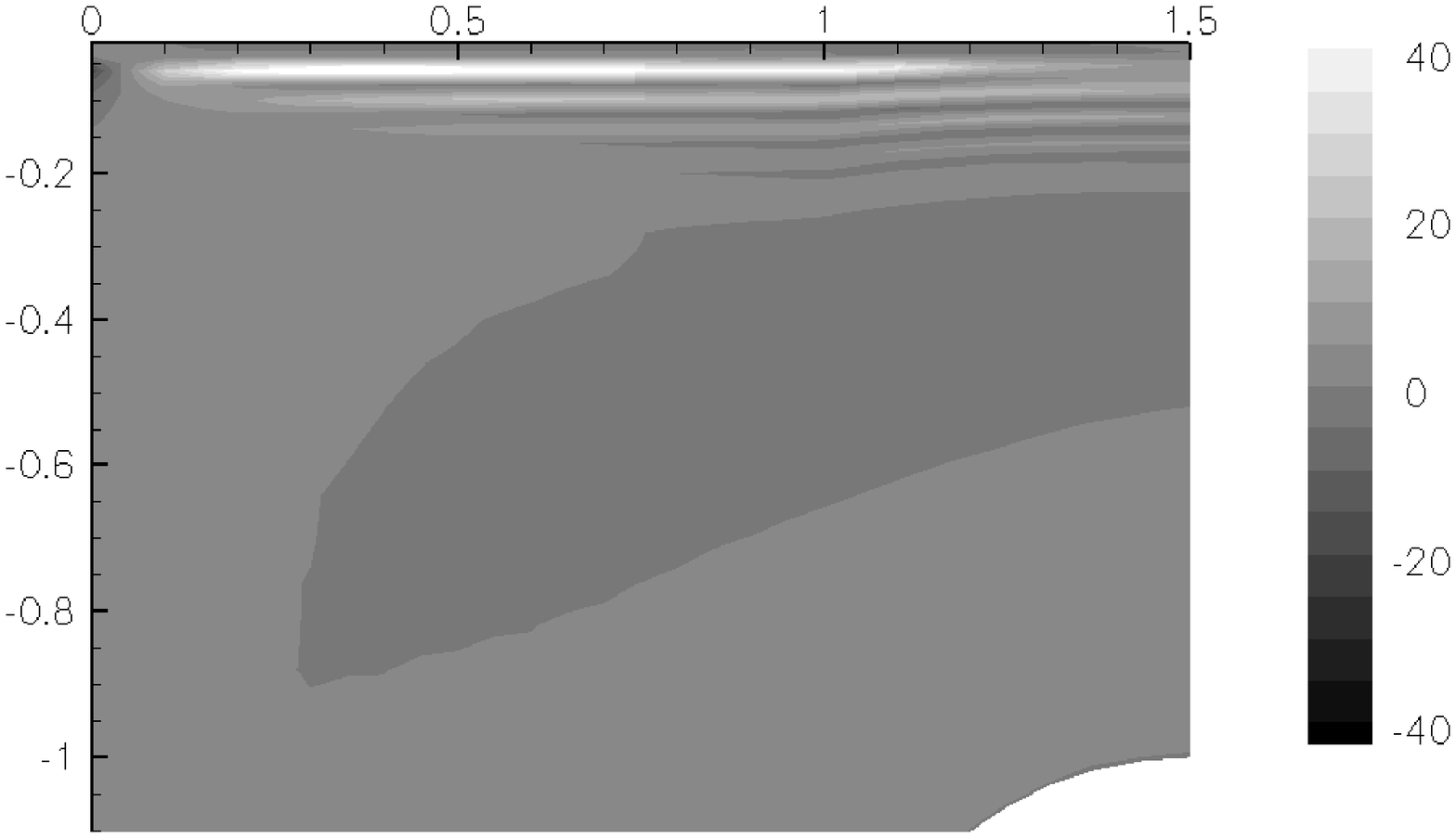}%
    }%
    \vspace{0pt}%
    \\%
    \subfigure[$~t = 3.35$]%
    {%
        \label{axx_c}%
        \includegraphics[width = 0.29\textheight, bb = 22 253 587 575]{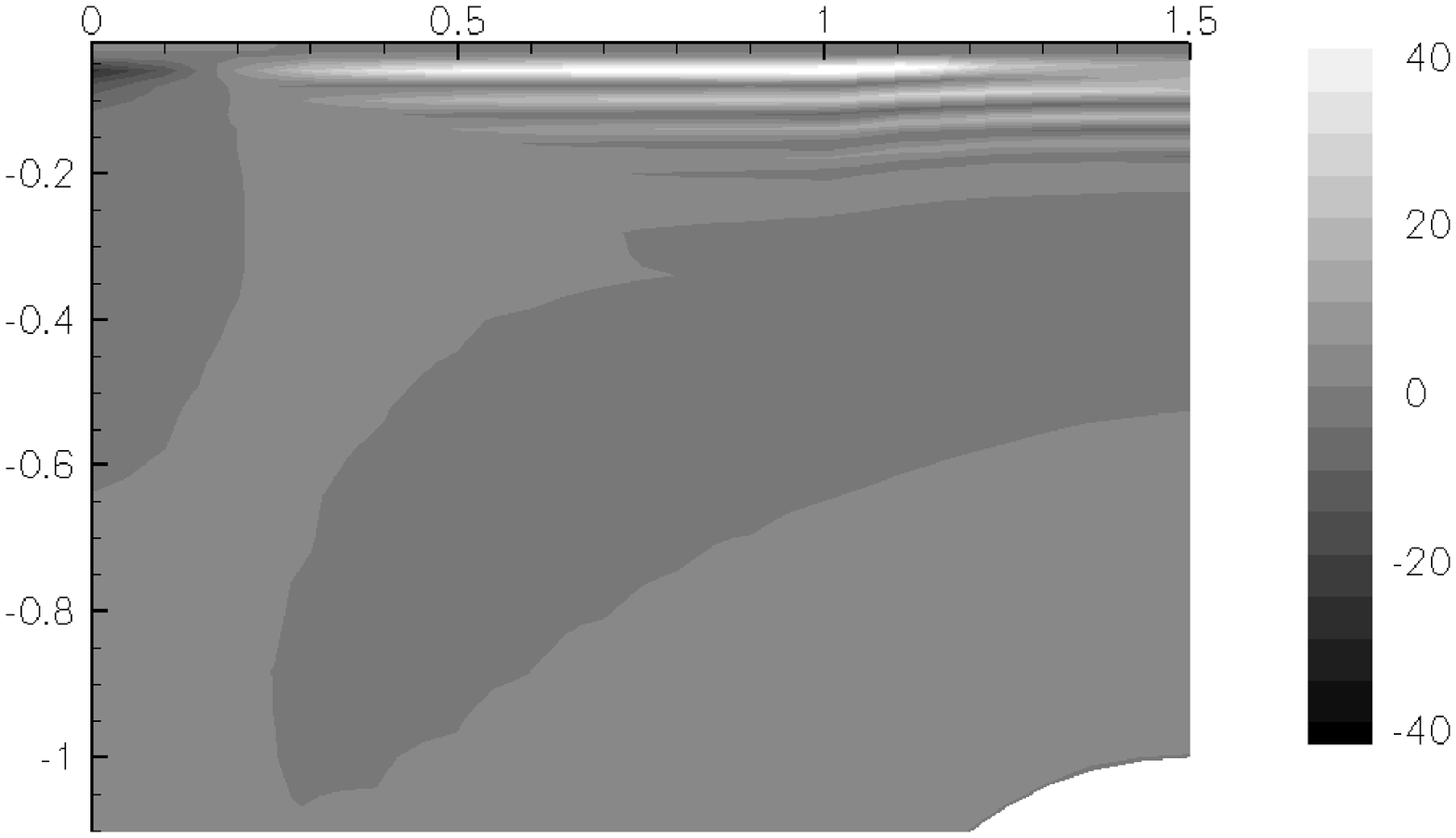}%
    }%
    \hspace{2pt}%
    \subfigure[$~t = 5.03$]%
    {%
        \label{axx_d}%
        \includegraphics[width = 0.29\textheight, bb = 22 253 587 575]{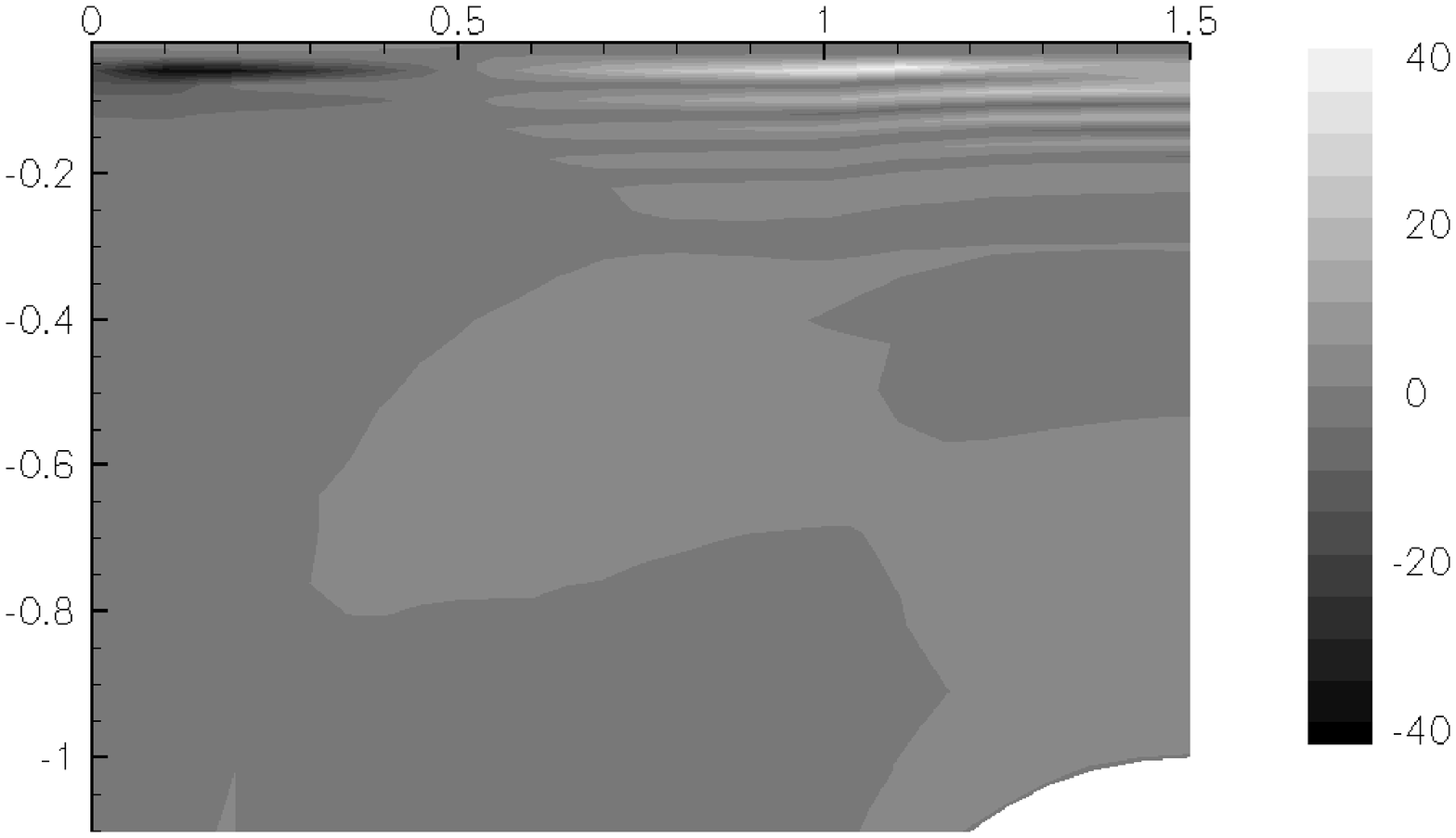}%
    }%
    \vspace{0pt}%
    \\%
    \subfigure[$~t = 6.71$]%
    {%
        \label{axx_e}%
        \includegraphics[width = 0.29\textheight, bb = 22 253 587 575]{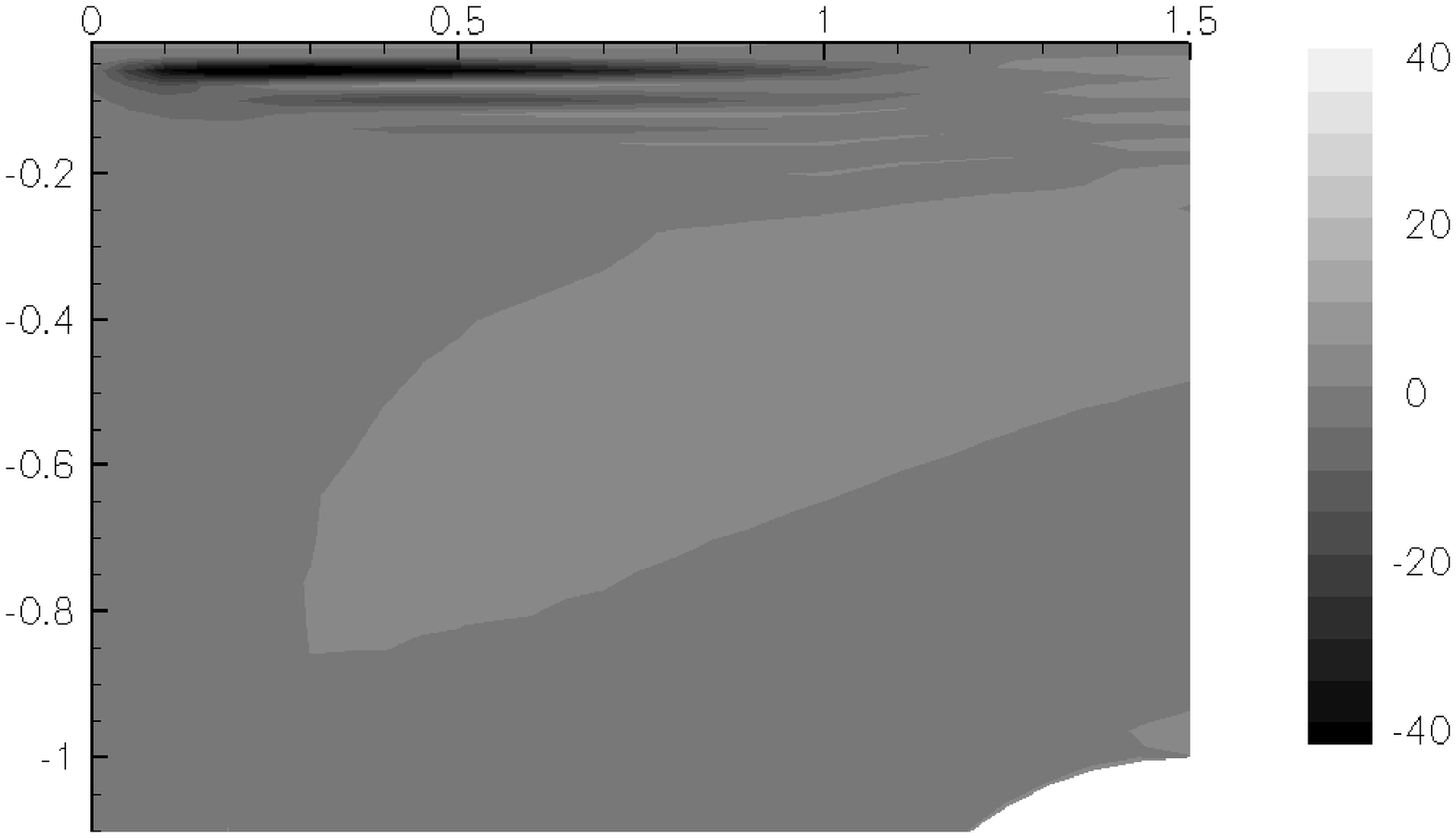}%
    }%
    \hspace{2pt}%
    \subfigure[$~t = 8.38$]%
    {%
        \label{axx_f}%
        \includegraphics[width = 0.29\textheight, bb = 22 253 587 575]{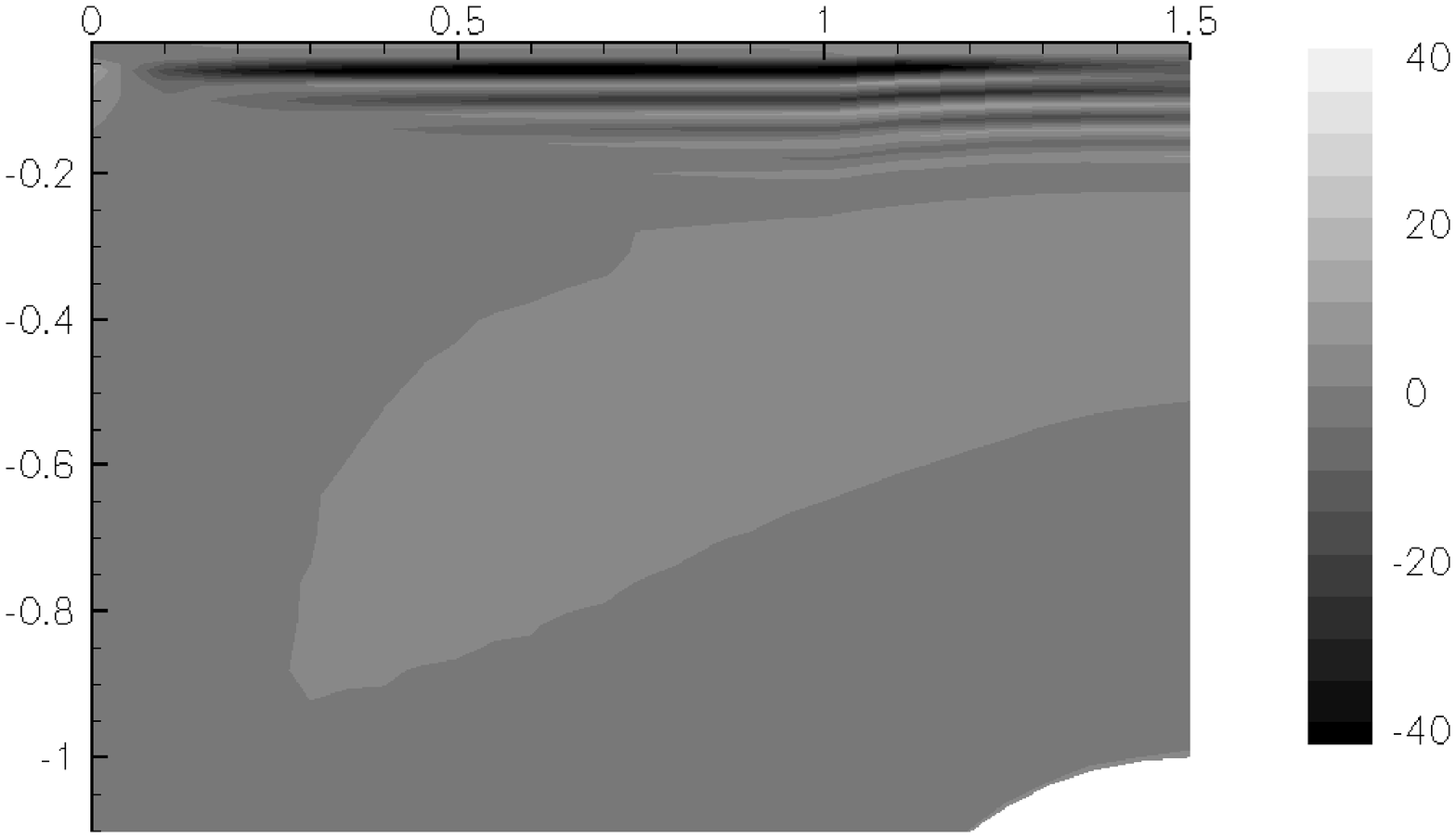}%
    }%
    \vspace{0pt}%
    \\%
    \subfigure[$~t = 10.06$]%
    {%
        \label{axx_g}%
        \includegraphics[width = 0.29\textheight, bb = 22 253 587 575]{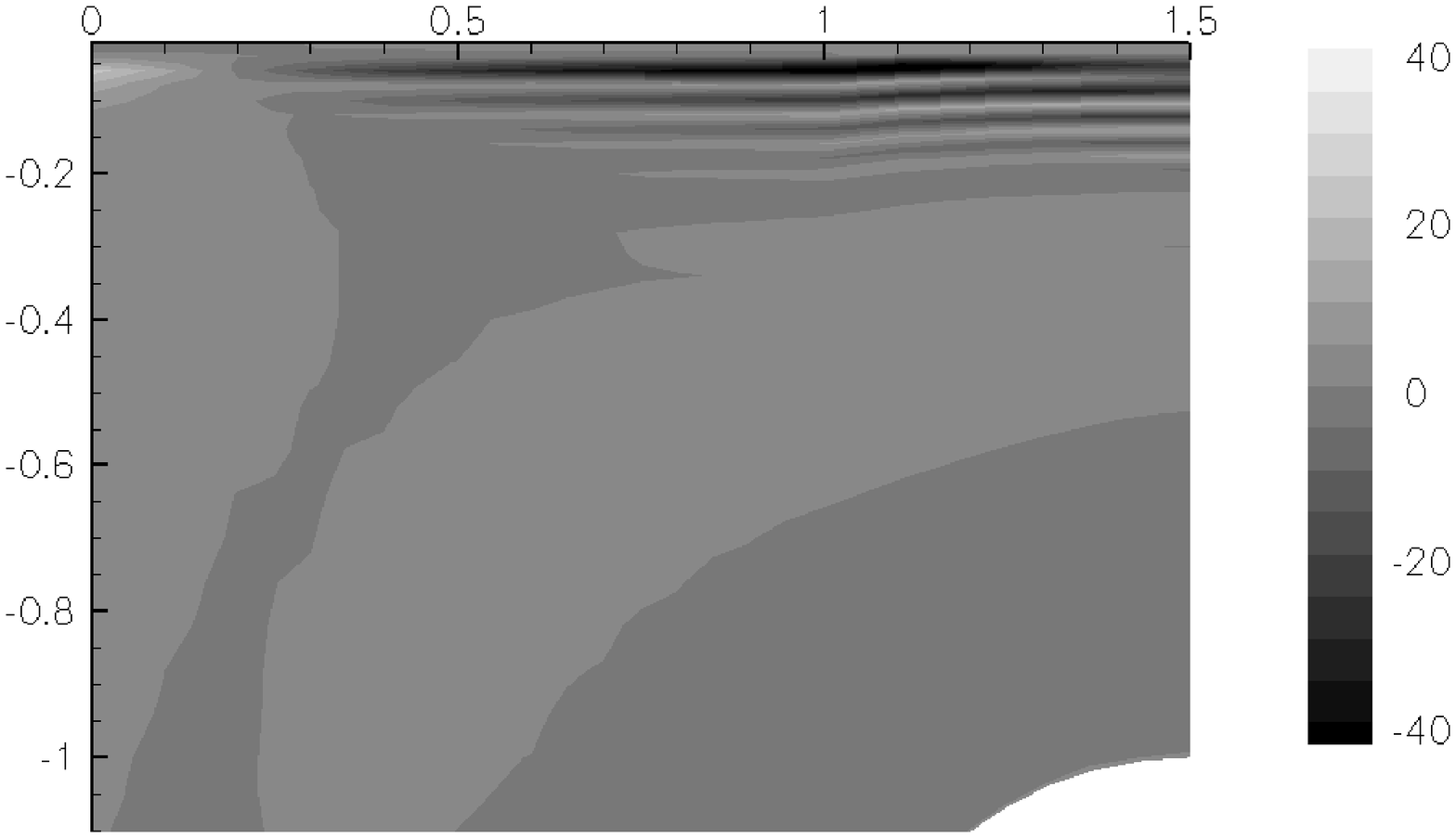}%
    }%
    \hspace{2pt}%
    \subfigure[$~t = 11.74$]%
    {%
        \label{axx_h}%
        \includegraphics[width = 0.29\textheight, bb = 22 253 587 575]{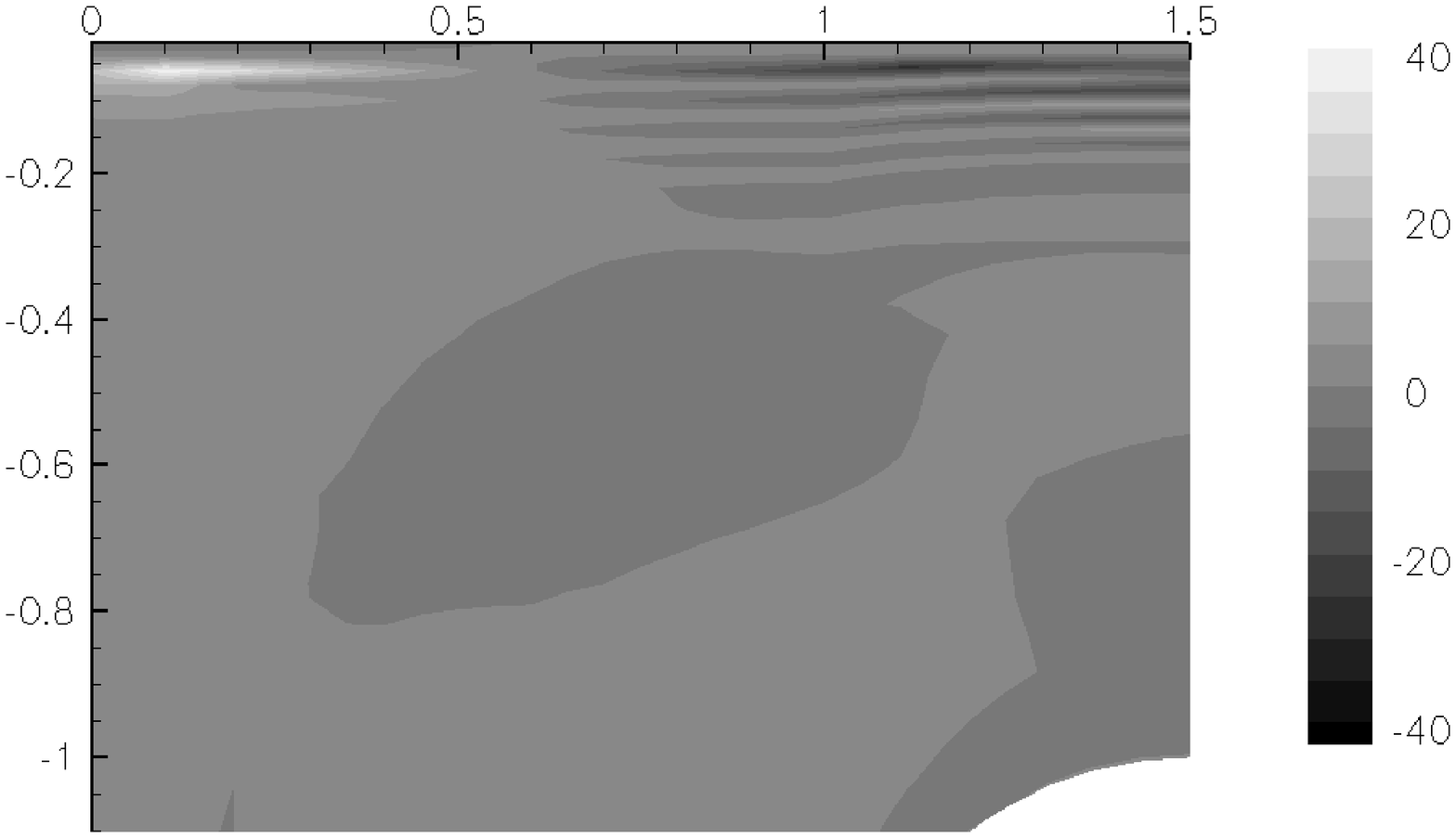}%
    }%
    \caption{Perturbation of xx-component of polymer conformation tensor, $\alpha_{xx}'$ with respect to steady states at periodic orbits: $\textit{Wi} = 66$. The region shown is $0<x<1.5$, $-1.1<y<0$, stagnation point is at the top-left corner. The edge of the steady state birefringent strand is the line $y\approx -0.05$.}%
    \label{axx_2006-11-5a}
\end{figure}

We turn now to the spatiotemporal structure of the instability and
its underlying physical mechanism. We will denote the deviations in
velocity, pressure and stress with primes, while steady state values
will be denoted with a superscript ``s'':
\begin{eqnarray}%
    \label{vel_decompose}%
        \boldsymbol{u} = \boldsymbol{u}^{\textrm{s}} + \boldsymbol{u}',%
    \\%
    \label{press_decompose}%
        p = p^{\textrm{s}} + p',%
    \\%
    \label{alpha_decompose}%
        \boldsymbol{\alpha} = \boldsymbol{\alpha}^{\textrm{s}} +
        \boldsymbol{\alpha}'.
\end{eqnarray}%
Figures~\ref{ux_2006-11-5a},~\ref{uy_2006-11-5a}~and~\ref{axx_2006-11-5a}
illustrate $u_{x}'$, $u_{y}'$ and $\alpha _{xx}'$, respectively, at
intervals of $1/8$ period, corresponding to the periodic orbit at a
Weissenberg number close to the bifurcation point ($\textit{Wi} =
66$). Time starts from an arbitrarily chosen snapshot on the
periodic orbit and only a quarter of the region near the stagnation
point is shown, behavior in the rest of the domain can be inferred
from the reflection symmetry across the axes.

At the beginning of the cycle~(Figure~\ref{ux_a}), $u_{x}'$ is
positive in the region very close to the stagnation point while it
is negative in most of the downstream region.
As time goes on, this positive deviation near the stagnation point
grows into a ``jet'', a region of liquid moving downstream away from
the stagnation point faster than the steady state velocity, as shown
in Figures~\ref{ux_b},~\ref{ux_c} and~\ref{ux_d}. Correspondingly,
by continuity, the inflow toward the stagnation point is also faster
as shown in Figures~\ref{uy_a}--\ref{uy_d}. Note that very near the
stagnation point deviations from steady state remain small. At the
beginning of the second half of the cycle (Figure~\ref{ux_e}), the
jet extends further downstream and grows to the full width of the
channel. Meanwhile, in the region closer to the stagnation point,
velocity deviations drop (Figures~\ref{ux_e},~\ref{uy_e}) and start
to change signs (Figures~\ref{ux_f},~\ref{uy_f}). Consequently, the
growth of the jet is interrupted and a ``wake'', a region of fluid
moving slower than the steady state velocity, emerges downstream
(Figures~\ref{ux_f}--~\ref{ux_h} and~\ref{uy_f}--~\ref{uy_h}).
Similarly, as the wake grow larger, velocity deviations near the
stagnation point change signs and a new cycle starts
(Figures~\ref{ux_a} and~\ref{uy_a}).

The velocity deviations are closely related with those of the stress
field (Figure~\ref{axx_2006-11-5a}). Generally speaking, ``jets''
are accompanied by negative $\alpha _{xx}'$ and thus thinning of the
birefringent strand and ``wakes'' are associated with the
birefringent thickening. The largest deviations are found at the
edges of the birefringent strand where $\partial\alpha
_{xx}^{s}/\partial y$ is largest. Note that deviations in the stress
field are always small along the centerline of the birefringent
strand because there polymer molecules are almost fully stretched
and the huge spring force is sufficient to resist any perturbations.

One may notice the small spatial oscillations in the stress field
deviations, characterized by alternating high and low stress
stripes, along the outflow direction. These oscillations, apparently
unphysical and centered around zero, also exist along the
birefringence strand in steady state solutions, though they are not
easy to see from the contour in Figure~\ref{tra_ss_2006-10-12g} as
they are overwhelmed by high $\textrm{tr}(\boldsymbol{\alpha})$ in
the birefringent strand. Unfortunately, as proven by
Renardy~\citep{Renardy_JNNFM2006}, spatial non-smoothness is
inevitable in numerical simulations of viscoelastic extensional flow
upon certain $\textit{Wi}$ due to the singularities in stress
gradients.
These singularities could not be fully resolved by any finite mesh
size and this problem would always show up in numerical solutions of
high $\textit{Wi}$ viscoelastic stagnation point flows. However, we
do not expect these oscillations to qualitatively affect our
observations for a couple of reasons. First, non-smoothness has been
observed in our simulation at $\textit{Wi}$ values much lower than
the critical $\textit{Wi}$ of this instability.
Second, observable non-smoothness is always found some distance away
from the stagnation point in the downstream direction while the
instability is dominated by the physics in the close vicinity of the
stagnation point and since FENE-P is a convective equation we do not
expect anything occurring downstream to affect upstream dynamics.
Last, and most importantly, simulations with different meshes
display different mesh size dependent stripes, while the nature of
the instability remains virtually unchanged.

\begin{figure}%
    \centering%
    \includegraphics[width = 0.8\textwidth]{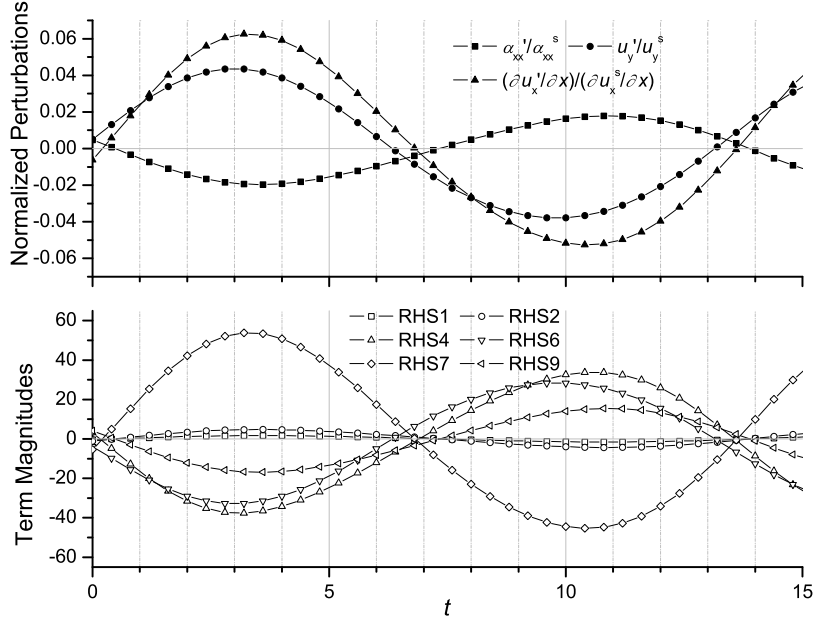}%
    \caption{Time-dependent oscillations at $(0, -0.05)$. Top view: perturbations of variables normalized by steady-state quantities; Bottom view: magnitudes of terms on RHS of Equation~(\ref{fenep_linear}).}
    \label{terms}%
\end{figure}

Insight into the mechanism of this instability can be gained by
examining the linearized equation for $\alpha _{xx}'$:
\begin{eqnarray}%
    \label{fenep_linear}%
\begin{split}%
        \frac{\partial\alpha _{xx}'}{\partial t} = & - \frac{2}{\textit{Wi}}\frac{\alpha _{xx}'}{1 -
        \frac{\textrm{tr}(\boldsymbol{\alpha}^{\textrm{s}})}{b}}%
        - \frac{2}{\textit{Wi}}\frac{\alpha _{xx}^{\textrm{s}}\textrm{tr}(\boldsymbol{\alpha}')}{b\left(1 -
        \frac{\textrm{tr}(\boldsymbol{\alpha}^{\textrm{s}})}{b}\right)^{2}}%
        \\%
        & - u_{x}^{\textrm{s}}\frac{\partial \alpha _{xx}'}{\partial x}%
        - u_{y}^{\textrm{s}}\frac{\partial \alpha _{xx}'}{\partial y}%
        - u_{x}'\frac{\partial \alpha _{xx}^{\textrm{s}}}{\partial x}%
        - u_{y}'\frac{\partial \alpha _{xx}^{\textrm{s}}}{\partial y}%
        \\%
        & + 2\alpha _{xx}^{\textrm{s}}\frac{\partial u_{x}'}{\partial x}%
        + 2\alpha _{xy}^{\textrm{s}}\frac{\partial u_{x}'}{\partial y}%
        + 2\alpha _{xx}'\frac{\partial u_{x}^{\textrm{s}}}{\partial x}%
        + 2\alpha _{xy}'\frac{\partial u_{x}^{\textrm{s}}}{\partial y}.%
\end{split}%
\end{eqnarray}%
In the following analysis, terms on the right-hand-side (RHS) of
Equation~\ref{fenep_linear} are named ``RHS$*$'', where ``$*$'' is
determined by the order of appearance on the RHS. Terms and their
physical meanings are summarized in Table~\ref{term_table}. To
understand the mechanism of the instability, magnitudes of these
terms at the point $(0, -0.05)$ are plotted as a function of time
during roughly a period in the bottom view of Figure~\ref{terms}.
Terms RHS3, RHS5, RHS8 and RHS10 are zero by symmetry and not
plotted. This position is right at the edge of the birefringent
strand and as shown in Figure~\ref{axx_2006-11-5a}, it is also where
significant deviations in the stress field are observed.
Time-dependent oscillations at other places, including off the
symmetry axis $x = 0$, have also been checked and nothing that could
qualitatively affect our analysis was seen. Correspondingly,
deviations in polymer conformation, inflow velocity and extension
rate, normalized by steady state values, are plotted in the top view
of Figure~\ref{terms}.

\begin{table}
\centering
\begin{minipage}{0.8\textwidth}
\centering%
\begin{tabular}{@{}l@{}l@{}l@{}}
    \hline \hline
    Term & Formula & Physical Significance
    \\ \hline \hline
     RHS1~ & $-\frac{2}{\textit{Wi}}\frac{\alpha _{xx}'}{1-\frac{\textrm{tr}(\boldsymbol{\alpha}^{\textrm{s}})}{b}}$
     & ~\parbox{0.65\textwidth}{Relaxation.}%
     \\ \hline
     RHS2~ & $-\frac{2}{\textit{Wi}}\frac{\alpha _{xx}^{\textrm{s}}\textrm{tr}(\boldsymbol{\alpha}')}{b\left(1-\frac{\textrm{tr}(\boldsymbol{\alpha}^{\textrm{s}})}{b}\right)^{2}}$
     & ~\parbox{0.65\textwidth}{Relaxation.}%
     \\ \hline
     RHS3~ & $-u_{x}^{\textrm{s}}\frac{\partial \alpha _{xx}'}{\partial x}$
     & ~\parbox{0.65\textwidth}{Convection of conformation deviations by the steady state x-velocity.}%
     \\ \hline
     RHS4~ & $-u_{y}^{\textrm{s}}\frac{\partial \alpha _{xx}'}{\partial y}$
     & ~\parbox{0.65\textwidth}{Convection of conformation deviations by the steady state y-velocity.}%
     \\ \hline
     RHS5~ & $-u_{x}'\frac{\partial \alpha _{xx}^{\textrm{s}}}{\partial x}$
     & ~\parbox{0.65\textwidth}{Convection of the steady state conformation by x-velocity deviations.}%
     \\ \hline
     RHS6~ & $-u_{y}'\frac{\partial \alpha _{xx}^{\textrm{s}}}{\partial y}$
     & ~\parbox{0.65\textwidth}{Convection of the steady state conformation by y-velocity deviations.}%
     \\ \hline
     RHS7~ & $2\alpha _{xx}^{\textrm{s}}\frac{\partial u_{x}'}{\partial x}$
     & ~\parbox{0.65\textwidth}{Stretching caused by deviations in the extension rate.}%
     \\ \hline
     RHS8~ & $2\alpha _{xy}^{\textrm{s}}\frac{\partial u_{x}'}{\partial y}$
     & ~\parbox{0.65\textwidth}{Stretching caused by deviations in the shear rate.}%
     \\ \hline
     RHS9~ & $2\alpha _{xx}'\frac{\partial u_{x}^{\textrm{s}}}{\partial x}$
     & ~\parbox{0.65\textwidth}{Stretching caused by deviations in the extensional stress.}%
     \\ \hline
     RHS10~ & $2\alpha _{xy}'\frac{\partial u_{x}^{\textrm{s}}}{\partial y}$
     & ~\parbox{0.65\textwidth}{Stretching caused by deviations in the shear stress.}%
     \\ \hline \hline
\end{tabular}
\caption{Terms on the right-hand side of Equation~(\ref{fenep_linear}).}%
\label{term_table}%
\end{minipage}
\end{table}

Consistent with our earlier observations, deviations in the velocity
field ($u_{y}'$ and $\partial u_{x}'/\partial x$) and deviations in
stress field ($\alpha _{xx}'$) are opposite in signs for most of the
time within the period. Among the terms plotted, RHS4, RHS6, RHS7
and RHS9 are much larger than the relaxation terms, RHS1 and RHS2,
and dominate the dynamics. (Relaxation terms are large at the very
inner regions of the birefringent strand and that is why
oscillations in the stress field there are barely noticeable.)
Moreover, RHS4, RHS6 and RHS9 are mostly in phase with $\alpha
_{xx}'$ and thus tend to enhance the deviations while RHS7 is out of
phase with $\alpha _{xx}'$ and hence damps the deviations. It is the
joint effect of these competing destabilizing and stabilizing forces
that gives the oscillatory behavior of the system. Finally, notice
that among the three destabilizing terms, RHS6 is the one that leads
the phase and thus guides the instability.

\begin{figure}
    \centering%
    \subfigure[~Thinning process of the birefringent strand.]%
    {%
        \label{schematic1}%
        \includegraphics[width = 0.7\textwidth, bb=0 0 792 397]{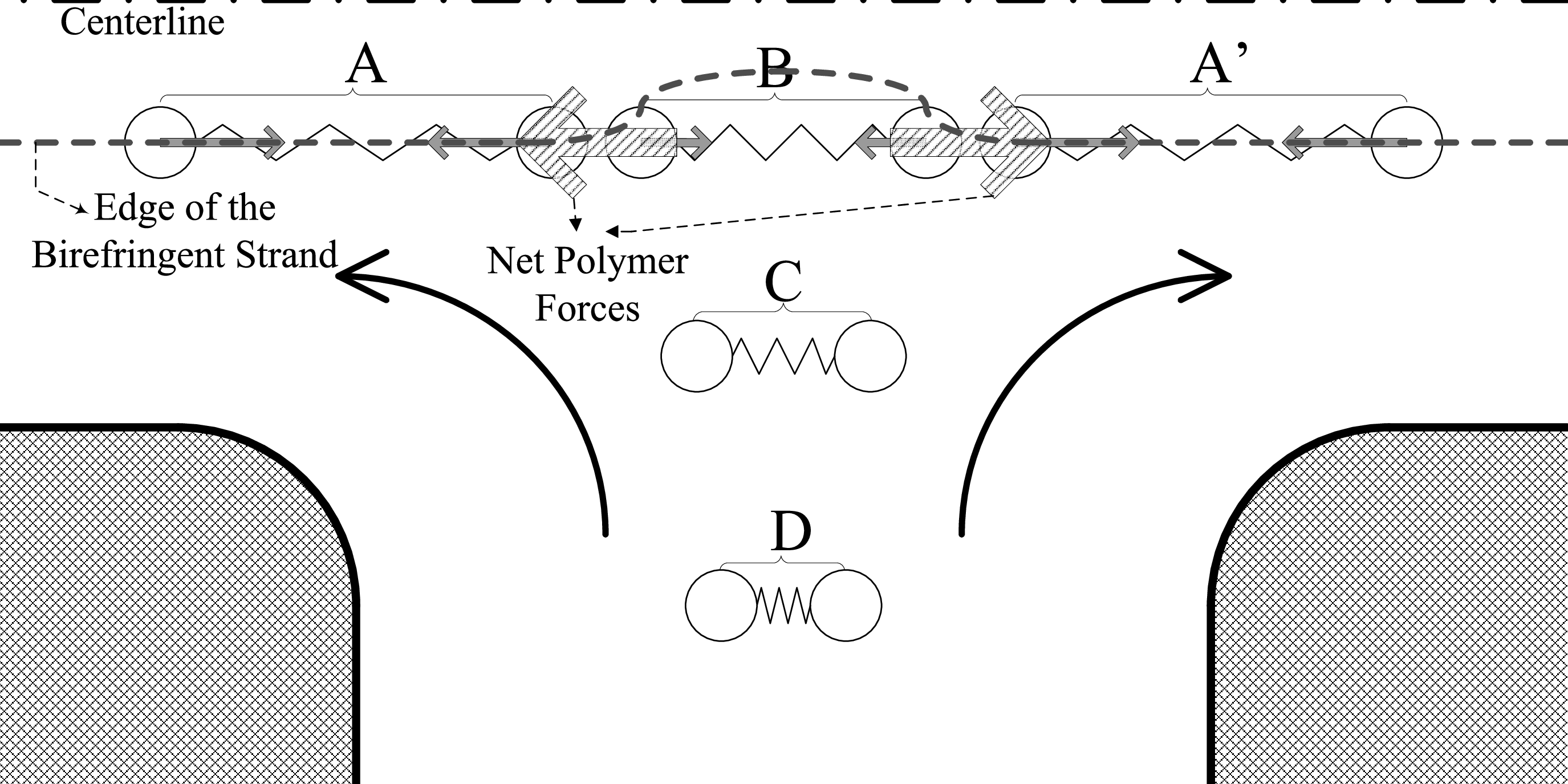}%
    }%
    \vspace{5pt}%
    \\%
    \subfigure[~Re-thickening process of the birefringent strand.]%
    {%
        \label{schematic2}%
        \includegraphics[width = 0.7\textwidth, bb=0 0 792 397]{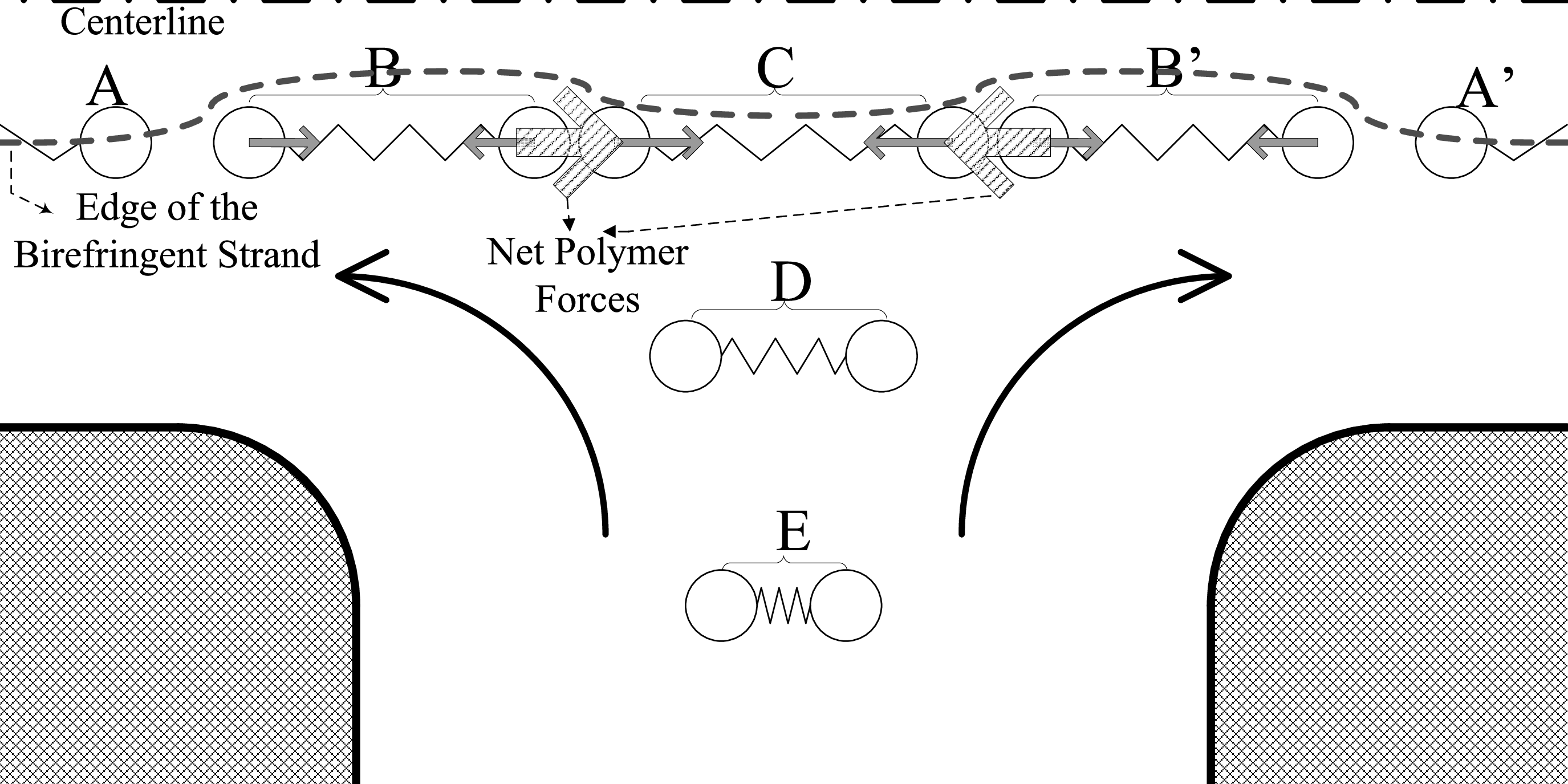}%
    }%
    \caption{Schematic of instability mechanism (view of the lower half geometry) (Gray solid arrows: spring forces of the dumbbells; Large shaded arrows: net forces exerted by polymer molecules (dumbbells) on the fluid).}%
    \label{schematic}%
\end{figure}

Based on these observations from Figure~\ref{terms}, a mechanism for
the instability can be proposed, which is illustrated schematically
in Figure~\ref{schematic}. At the beginning of the cycle ($t = 0$),
$u_{y}'$ is slightly above zero, indicating that the inflow speed is
faster than that in the steady state. As a consequence, RHS6 becomes
negative first, followed by RHS4 and RHS9. In particular, a faster
incoming convective flow brings unstretched polymer molecules toward
the stagnation point (corresponding to RHS6), as depicted in
Figure~\ref{schematic1}.
These polymer chains have less time to get stretched and when they
reach the edges of the birefringent strand (e.g. dumbbell B), they
are less stretched compared with the steady state. As a result,
fluid around dumbbell B has lower stress than at the steady state,
corresponding to a thinning of the birefringent strand. Meanwhile,
since dumbbell B contains smaller spring forces than its downstream
neighbors A and A', the net forces (shaded arrows) exerted by
polymer on the fluid point outward, generating jets downstream from
the stagnation point. (In other words, when the stress at the center
is lower, the net stress divergence points outward, which increases
momentum in the downstream directions.) By continuity, more fluid
has to be drawn toward the stagnation point and the initial
deviation in $u_{y}'$ is then enhanced. However, as the flow speeds
up in the vicinity of the stagnation point, the extension rate also
starts to increase. This effect (corresponding to RHS7) tends to
stretch polymer molecules more and stabilize the deviations, as
shown in Figure~\ref{terms}.
Eventually this effect will be able to overcome that of RHS6 as well
as RHS4 and RHS9 and the stress near the stagnation point starts
increase after it passes the minimum at around $t = 3.5$, which
causes a re-thickening of the birefringent strand as illustrated in
Figure~\ref{schematic2}. By a similar argument as that above,
dumbbell C has higher spring forces than B and B', the dumbbells
which were passing near the center when stress was at minimum, and
the net polymer forces point inward, which starts to suppress the
jets. Inflow velocity decreases as the birefringent strand thickens
and this gives incoming polymer molecules more time to be stretched
and further thickens the birefringent strand. Eventually $\alpha
_{xx}$ will come back to the steady state value at around $t = 7.2$,
however, since all the deviations are not synchronized, a negative
deviation is found in $u_{y}$ and an identical analysis with
opposite signs can be made for the second half of the cycle.

Within this mechanism, a sharp edge of the birefringent strand, i.e.
large magnitude of $\partial\alpha _{xx}/\partial y$ ($\sim \mathcal
{O}(10^{4})$ in our simulations), is required so that a small
$u_{y}'$ can give a sufficiently large RHS6 to drive the
instability. This is made possible by the kinematics of the flow
near the stagnation point, where the incoming polymer molecules are
strongly stretched within a short distance. Another similar effect
is that stress derivatives are stretched in the outgoing direction
and thus greatly weakened as fluid moves downstream; therefore the
instability is dominated by physics in the vicinity of the
stagnation point. In the earlier mechanism for the so-called
``varicose instability'', given by Harris and
Rallison~\citep{Harris_JNNFM1994}, the importance of extensional
stress and flow kinematics, especially the role of convection of
incoming molecules, was also recognized. However, the picture
described in their work is not the same as ours due to the
simplifications in their model. Their linear stability analysis
ignores the x-dependence of the birefringent width while in our
simulations, x-dependence of the stress field is closely related to
the changes in velocity field.
Besides, their analysis does not identify a restoring force for the
deviations and the oscillatory behavior could not be explained.

\section{Conclusions}
Using a DEVSS/SUPG formulation of the finite element method, we are
able to simulate viscoelastic stagnation point flow and obtain
steady state and time-dependent solutions at high $\textit{Wi}$. For
$\textit{Wi}\gg 1$, a clear birefringent strand is observed.
The width of this birefringent strand increases with increasing
$\textit{Wi}$ until $\textit{Wi}\approx 40$ after which it declines
gradually. This also results in a non-monotonic trend in the
modification of the velocity field.

At around $\textit{Wi} = 65$ the steady state solution loses
stability and a periodic orbit becomes the attractor in phase space.
Flow motion of the periodic orbit is characterized by time-dependent
fluctuations, specifically, alternating positive (jet) and negative
(wake) deviations from the steady state velocity in the regions
downstream of the stagnation point.
A mechanism is proposed which, taking account of the interaction
between velocity and stress fields, is able to explain the whole
process of the oscillatory instability. Extensional stresses and
their gradients as well as the flow kinetics near the stagnation
points is identified as important factors in the mechanism. This
mechanism is different from that of the ``hoop stress''
instabilities, which occur in viscometric flows with curved
streamlines, and we expect that this mechanism could be extended and
explain various instabilities occurring in viscoelastic flows with
stagnation points.

\section{Acknowledgement}
The authors would like to acknowledge financial support for this
research from the National Science Foundation and the Petroleum
Research Fund, administered by the American Chemical Society.

\bibliographystyle{jfm}
\bibliography{StagPt}

\end{document}